\documentclass[twocolumn]{aastex62}
\pdfoutput=1 
\usepackage{hyperref}
\usepackage{amsmath,amstext}
\usepackage[T1]{fontenc}
\usepackage{apjfonts} 

\usepackage[hang,flushmargin]{footmisc} 

\usepackage[nameinlink,capitalize]{cleveref}
\crefname{section}{section}{section}
\Crefname{Section}{Section}{Section}

\usepackage{url}

\usepackage{graphicx}
\usepackage[figure,figure*]{hypcap}
\usepackage[caption=false]{subfig}

\graphicspath{{./figs/}}

\hypersetup{
    allcolors = {blue},
}

\bibpunct{(}{)}{;}{a}{}{,}

\def\msun{{\rm\,M_\odot}}

\newcommand{\cspg}{\rm\ cm^2\ g^{-1}}
\newcommand{\ergs}{\rm\ erg\ s^{-1}}

\newcommand{\mosfit}{{\tt MOSFiT}}

\definecolor{orange}{rgb}{.8,0.4,0}

\newcommand{\Mej}{M_{\rm ej}}
\newcommand{\vej}{v_{\rm ej}}
\newcommand{\apx}{\ensuremath{\sim}}

\defcitealias{LP98}{LP98}
\defcitealias{CowpBerger15}{CB15}

\shorttitle{Kilonova Observations With LSST}
\shortauthors{Cowperthwaite et~al.}

\begin{document}

\title{LSST Target-of-Opportunity Observations of Gravitational Wave Events: Essential and Efficient}

\author[0000-0002-2478-6939]{P.~S.~Cowperthwaite}
\affiliation{The Observatories of the Carnegie Institution for Science, 813 Santa Barbara St., Pasadena,
CA 91101, USA}
\altaffiliation{Hubble Fellow, \href{mailto:pcowperthwaite@carnegiescience.edu}{pcowperthwaite@carnegiescience.edu}}
\author[0000-0002-5814-4061]{V.~A.~Villar}
\affiliation{Harvard-Smithsonian Center for Astrophysics, 60 Garden Street, Cambridge, Massachusetts 02138, USA}
\author[0000-0002-4934-5849]{D.~M.~Scolnic}
\affiliation{Kavli Institute for Cosmological Physics, The University of Chicago, Chicago, IL 60637}
\author[0000-0002-9392-9681]{E.~Berger}
\affiliation{Harvard-Smithsonian Center for Astrophysics, 60 Garden Street, Cambridge, Massachusetts 02138, USA}

\begin{abstract}
We present simulated observations to assess the ability of LSST and the WFD survey to detect and characterize kilonovae -- the optical emission associated with binary neutron star (and possibly black hole -- neutron star) mergers. We expand on previous studies in several critical ways by exploring a range of kilonova models and several choices of cadence, as well as by evaluating the information content of the resulting light curves. We find that, depending on the precise choice of cadence, the WFD survey will achieve an average kilonova detection efficiency of $\approx 1.6-2.5\%$ and detect only $\approx 3-6$ kilonovae per year. The detected kilonovae will be within the detection volume of Advanced LIGO/Virgo (ALV).  By refitting the best resulting LSST light curves with the same model used to generate them we find the model parameters are generally weakly constrained, and are accurate to at best a factor of $2-3$.  Motivated by the finding that the WFD will yield a small number of kilonova detections, with poor light curves and marginal information content, and that the detections are in any case inside the ALV volume, we argue that target-of-opportunity follow-up of gravitational wave triggers is a much more effective approach for kilonova studies.  We outline the qualitative foundation for such a program with the goal of minimizing the impact on LSST operations. We argue that observations in the $gz$-bands with a total time investment per event of $\approx 1.5$ hour per 10 deg$^2$ of search area is sufficient to rapidly detect and identify kilonovae with $\gtrsim 90\%$ efficiency. For an estimated event rate of $\sim20$ per year visible to LSST, this accounts for $\sim1.5\%$ of the total survey time. In this regime, LSST has the potential to be a powerful tool for kilonovae {\it discovery}, with detected events handed off to other narrow-field facilities for further monitoring.
\end{abstract}


\section{Introduction}
\label{sec:intro}
The discovery of an optical counterpart associated with the binary neutron star merger GW170817 was a watershed moment in the development of joint gravitational wave and electromagnetic (GW-EM) astronomy \citep{LIGOGW170817,LIGOMMAPaper,Arcavi+17,Coulter+17,GW170817DECam,Valenti+17}. Modeling of the resulting UV/optical/NIR light curves revealed behavior consistent with that of a kilonova \citep{Cowp+17,Drout+17,Kilpatrick+17,Tanaka+17,Villar+17b, Tanaka+18, Villar+18}, a transient powered by radioactive decay of $r$-process material synthesized in the ejecta from mergers of compact object binaries involving at least one neutron star \citep[see][for a review]{Metzger2017}. The identification of an optical counterpart also enabled a range of other critical studies, including: extensive optical and NIR spectroscopic observations \citep{Chornock+17,Nicholl+17a,Shappee+17}, the discovery of radio and X-ray emission which demonstrated the presence of an off-axis jet \citep{Alexander+17,Hallinan+17,Margutti+17,Troja+17,Lazzati+18,Margutti+18,Mooley+18}, studies of the host galaxy \citep[NGC4993, see e.g.,][]{Blanchard+17,Cantiello+18}, joint GW-EM constraints of the neutron star equation of state \citep[see e.g.,][]{De+18,LIGONSEOS,Radice+18}, and independent measurements of local cosmological parameters \citep[H$_0$, see e.g.,][]{LIGOH0,Guidorzi+17}.

Building upon the success of GW170817 and expanding GW-EM science relies on on-going improvements in the GW detector network. Over the next few years Advanced LIGO and Virgo (ALV) will reach their design sensitivity, while additional interferometers, KAGRA in Japan \citep{KAGRA} and LIGO-India \citep{LIGOIndia}, will join the network. In this multi-detector regime, binary neutron star mergers will be detected to $\apx200$~Mpc (and several hundred Mpc for NS-BH mergers) and  localized to $\apx10$~deg$^2$ \citep{Fairhurst2014,ChenHolz16}.

Similarly, we expect that at the larger detection distances next generation optical facilities may be crucial for counterpart identification. One such facility, the Large Synoptic Survey Telescope \citep[LSST,][]{Ivezic+09}, will be the premiere time-domain instrument in the Southern hemisphere during the next decade. LSST, with an 8.4-m primary mirror and a 9.6 deg$^2$ field-of-view, is well suited to the task of gravitational wave follow-up. 

Maximizing the potential of LSST requires the development of efficient and effective strategies for gravitational wave follow-up. This includes an investigation of both kilonovae detected in the LSST main survey (with and without associated GW signals), as well as via target-of-opportunity follow-up of GW-triggered events. \citet{Scolnic+18} addressed the first point by injecting model light curves of the kilonova associated with GW170817 from \citet{Cowp+17} into the LSST wide-fast-deep (WFD) survey using the current cadence ({\tt minion\_1016}). They found that LSST should detect $\apx7$ GW170817-like kilonovae per year. However, these kilonovae will have only $4-5$ data points per event, leading to poorly-sampled light curves and delayed identification. As a result, identification and modeling of these kilonovae will be challenging. Furthermore, only $\apx15\%$ of these kilonovae are expected to be detected within the detection volume of the GW network at design sensitivity, indicating that GW detections of these kilonovae will be unlikely. 

This led \citet{Scolnic+18} to suggest that target-of-opportunity observations of GW-triggered events are a more promising approach. A similar conclusion was reached in a model-agnostic study of the probability that baseline LSST operations will observe a GW localization region by pure chance, finding that the likelihood was vanishingly small \citep[see Chapter 6.5;][]{LSSTScienceBook}.

Here we expand on the work of \citet{Scolnic+18} in several critical ways. First, we increase the variety of kilonovae model light curves injected into the LSST WFD survey with several choices of observing cadence. This is accomplished by constructing models from a wider range of ejecta masses and compositions to fully explore the potential range of kilonovae brightness, timescale, and colors. Second, we model the resulting light curves from the WFD survey to assess their information content. Lastly, we discuss the design and benefits of target-of-opportunity observations triggered by the detection of a GW event.

This paper is organized as follows: In \autoref{sec:models} we describe the kilonova models used in our simulated observations. In \autoref{sec:obs} we describe our methodology for simulating LSST observations in the context of three different cadences using the SNANA analysis package. In \autoref{sec:WFD_results} we present the results of our simulated WFD survey observations of kilonovae. In \autoref{sec:too_results}, we discuss target-of-opportunity strategies with LSST. We conclude with a summary of the key results  in \autoref{sec:conc}.

All magnitudes presented in this work are given in the AB system. Cosmological calculations are performed using the cosmological parameters $H_0 = 67.7$ km s$^{-1}$ Mpc$^{-1}$, $\Omega_M = 0.307$, and $\Omega_{\Lambda} = 0.691$ \citep{Planck2016}.

\section{Kilonova Models \label{sec:models}}

A kilonova is an isotropic UV/optical/NIR (UVOIR) thermal transient powered by radioactive decay of $r$-process material synthesized during the merger of a compact object binary containing at least one neutron star. In this scenario, the merger is expected to produce a small amount of ejecta ($\Mej \lesssim 0.1 \msun$), which is typically neutron-rich \citep{Hotokezaka+13,Kyutoku+15,Metzger2017,SiegelMetzger17}. Heavy nuclei synthesized as the ejecta expands and decompresses are unstable and will decay on a wide range of timescales depositing energy into the ejecta which powers the kilonova emission \citep{LP98,Metzger+10,BarnesKasen13,TanakaHotokezaka13,Metzger2017}.

The exact nature of the kilonova emission depends strongly on the composition of the ejecta, specifically the neutron-richness which is parametrized by the electron fraction ($Y_e$). If the material is very neutron-rich ($Y_e \lesssim 0.25$), then the ejecta will undergo strong $r$-process nucleosynthesis, producing heavy elements ($A > 140$), particularly those in the lanthanide and actinide series. As a result, the ejecta will have a high opacity $(\kappa \gtrsim 10 \cspg)$. The resulting ``red" kilonova will then be faint; with a peak bolometric luminosity of $L_p \apx10^{40}-10^{41} \ergs$, red ($i-z \gtrsim 0.3$), and exhibit a timescale of $t_p \apx1$ week \citep{BarnesKasen13,TanakaHotokezaka13}. If instead the material is less neutron-rich ($Y_e \gtrsim 0.25$), then the ejecta will undergo light $r$-process nucleosynthesis, producing Fe-group and light $r$-process elements ($A \lesssim 140$), and the ejecta will have a lower opacity $(\kappa \apx0.1 \cspg)$. The resulting ``blue" kilonova will be brighter $L_p \apx10^{41}-10^{42} \ergs$, bluer ($i-z \lesssim 0$), and exhibit a shorter timescale of $t_p \apx1$ day \citep{Metzger+10,MetzgerFernandez14}.

\begin{figure*}[!t]
\label{fig:lc_full}
\begin{center}
\hspace*{-0.1in}
\scalebox{1.}
{\includegraphics[width=0.75\textwidth]{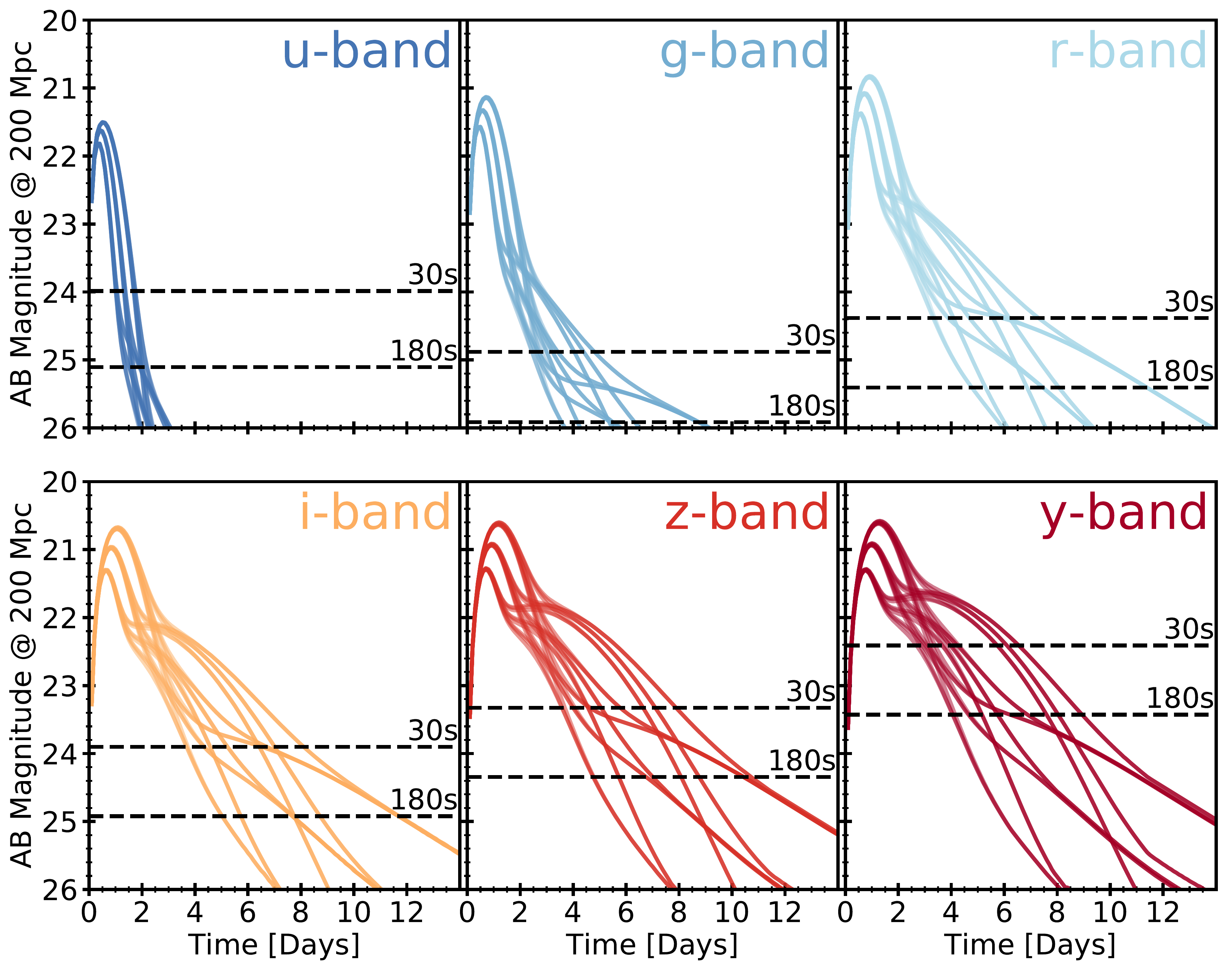}}
\caption{Light curves computed from the suite of kilonova models described in \autoref{sec:models}. Light curves are shown in all six LSST filters and scaled to a fiducial distance of $200$~Mpc. The dashed lines indicate the $5\sigma$ limiting magnitudes for integration times of $30$~s and $180$~s. These limiting magnitudes are computing assuming ideal observing conditions.}
\end{center}
\end{figure*}

In practice, the observed kilonova will exhibit some combination of both ``red" and ``blue" features. This behavior was seen in the kilonova associated with GW170817, where the multi-band UVOIR light curve was best described by a three-component model consisting of ``red"  $(\kappa \apx10 \cspg)$, ``purple"  $(\kappa \apx3 \cspg)$, and "blue"  $(\kappa \apx0.1 \cspg)$ $r$-process powered components \citep{Cowp+17,Villar+17b}. This three-component approach was first suggested by \citet{Tanaka+17} as a good approximation to the more complex opacity behavior seen in detailed radiative transport simulations. The three-component approach is also physically motivated as the different emission components may arise from different sources of ejecta present in the merger. The ``blue`` emission may arise from neutron-poor ejecta in the polar region where material is shock-heated during the NS collision \citep{Oechslin+07,Bauswein+13a,Sekiguchi+16} or if the ejecta is irradiated by neutrinos from a long-lived merger remnant \citep{FernandezMetzger13,Just+15,Kasen+15}. The ``purple" and ``red" emission may arise from ejecta produced in the tidal tails during merger \citep{Rosswog+99,Hotokezaka+13} or wind outflows from a post-merger accretion disk \citep{Just+15,SiegelMetzger17}. However, it is currently unknown if the behavior observed in GW170817 is ubiquitous, and whether it depends on parameters such as the mass ratio, the nature of the binary, and viewing orientation.. For example, if the ``blue" emission is indeed constrained to the polar regions of the ejecta then it may not be observable for all viewing angles, whereas the more isotropic ``red" emission from tidal tails or post-merger disks will be ubiquitous.

\begin{deluxetable*}{lcccccc}
\tabletypesize{\footnotesize}
\tablecolumns{7}
\tablewidth{0pt}
\tablecaption{Summary of WFD Survey Kilonova Detection Efficiencies and Rates \label{tab:rates}}	
\tablehead{
    \colhead{Cadence} &
    \colhead{$\epsilon_{\rm det} \; (N_{\rm det})$} &  
    \colhead{$\epsilon_{\rm rise} \; (N_{\rm rise})$} & 
    \colhead{$\epsilon_{\rm color} \; (N_{\rm color})$} &
    \colhead{$\epsilon_{\rm both} \; (N_{\rm both})$} &
    \colhead{$\epsilon_{< 200~{\rm Mpc}} \; (N_{< 200~{\rm Mpc}})$} &
    \colhead{$\epsilon_{< 450~{\rm Mpc}} \; (N_{< 450~{\rm Mpc}})$} \\
    \colhead{ } &
    \colhead{[$- \; ({\rm yr}^{-1})$]} & 
    \colhead{[$- \; ({\rm yr}^{-1})$]} &
    \colhead{[$- \; ({\rm yr}^{-1})$]} &
    \colhead{[$- \; ({\rm yr}^{-1})$]} &
    \colhead{[$- \; ({\rm yr}^{-1})$]} &
    \colhead{[$- \; ({\rm yr}^{-1})$]}
}
\startdata
WFDM & $1.6\% \; (3.7)$ & $0.8\% \; (1.7)$ & $0.3\% \; (0.8)$ & 
       $0.2\% \; (0.4)$ & $14\% \; (1.9)$ & $6.8\% \; (3.5)$ \\
WFDD & $2.5\% \; (6.6)$ & $1.6\% \; (3.4)$ & $2.4\% \; (6.4)$ & 
       $1.6\% \; (3.3)$ & $22\% \; (3.4)$ & $11\% \; (6.5)$ \\
WFDR & $1.0\% \; (3.0)$ & $0.6\% \; (1.5)$ & $0.4\% \; (1.3)$ & 
       $0.3\% \; (0.6)$ & $8.5\% \; (1.3)$ & $4.3\% \; (2.9)$ \\ 
\enddata
\tablecomments{Summary of the expected kilonova detection efficiencies and rates for each choice of WFD survey cadence and each choice of light curve criteria (see \autoref{sec:WFD_criteria}). We also compute the expected rates within the average and maximum ALV BNS sensitivity range, finding that the majority of kilonova expected to be detected by LSST will also be found with GW observations.}
\end{deluxetable*}

We explore this potential diversity in the context of LSST observations by producing a grid of models that cover a range of ejecta parameters. We generate synthetic spectral energy distributions (SEDs) for kilonovae spanning $0.2-1.5$ \micron\ using the {\tt kilonova} model built into the \mosfit\ light curve fitting package \citep{Guillochon+17b,Nicholl+17b}. The {\tt kilonova} model is discussed in \citet{Villar+17b} and describes a three-component kilonova \citep{Metzger2017,Villar+17b}. Each component is described by the {\tt rprocess} model \citep[see e.g.,][]{Villar+17a} parametrized by the ejecta mass ($\Mej$), velocity ($\vej$), and gray opacity ($\kappa$). 

We produce three-component kilonova models consisting of ``blue,'' ``purple,'' and ``red'' components. For each component, we make three physically motivated choices of possible ejecta masses starting from the best fit posteriors for GW170817 from \citet{Villar+17b} and extrapolating beyond GW170817 to a low, moderate, and high amount of ejected material. The parameter ranges are: $\rm\,M_{\rm ej,blue} = [0.005, 0.01, 0.02]$ $\msun$, $\rm\,M_{\rm ej,purple} = [0.01,0.02,0.05]$ $\msun$, and $\rm\,M_{\rm ej,red} = [0.005, 0.01, 0.02]$ $\msun$. We select values for $\vej$ based on modeling of GW170817 and expectations from simulations \citep{Hotokezaka+13,Kyutoku+15,Barnes+16,Cowp+17,Metzger2017,Villar+17b}, namely $v_{\rm ej,blue}=0.25$c and $v_{\rm ej,red/purple}=0.15$c. Similarly, following GW170817 for physical guidance, we fix the SED temperature floors at $T_{\rm floor, blue} = 800$~K, $T_{\rm floor, purple} = 1250$~K, and $T_{\rm floor, red} = 3800$~K. Lastly, the grey opacities are fixed at $\kappa_{\rm blue} = 0.5 \cspg$, $\kappa_{\rm purple} = 3 \cspg$, and $\kappa_{\rm red} = 10 \cspg$. We combine all permutations of components and ejecta masses to produce a final set of twenty-seven model light curves that cover a range of brightnesses, timescales, and colors. These model light curves in the LSST bands ($ugrizY$) are shown in \autoref{fig:lc_full}.

\section{Simulating LSST Observations}
\label{sec:obs}
\subsection{WFD Survey Observations}
\label{sec:obs_main}
We produce simulated observations of the 27 kilonova model permutations in the LSST WFD survey using the SNANA simulation package \citep{SNANA}. SNANA produces synthetic light curves using the kilonova models described in \autoref{sec:models} and a cadence library that contains a list of observations and observing conditions such as the photometric zero point, sky noise, and point spread function (PSF). For the LSST WFD survey, we produce this cadence library using the LSST Operations Simulator \citep[OpSim,][]{OpSim1,OpSim2}. This is a publicly available package designed to produce realistic simulations of LSST scheduling and imaging over the ten year duration of the survey. These simulations provide realistic information about observations including cadence, observing conditions due to telescope or environmental factors, and image characteristics.

In addition to the most recent OpSim reference run ({\tt minion\_1016}, hereafter WFDM), we explore two additional choices of cadence for the WFD survey: (i) an alternative scheduler (hereafter WFDD\footnote{\url{http://altsched.rothchild.me:8080/}}) which scans along with the meridian using scheduled blocks rather than a greedy optimizer, and increases the number of nightly filter changes; and (ii) a rolling cadence (hereafter WFDR), which toggles between rapid and slow cadences across different sections of the sky for different years and is optimized over certain features including $5\sigma$ limiting depth, target goal map, and slewtime.

We inject 200,000 model light curves into each of the three choices of cadence. For each injection, we uniformly choose a random model from the set of twenty-seven described in \autoref{sec:models}. The chosen model is injected at a random sky position in the LSST WFD survey footprint and at a random time, both chosen uniformly. The model is injected at a random distance out to a maximum of $D_{\rm max} \approx 1$~Gpc $(z\approx0.2$). The distance is chosen uniformly to ensure there are enough sources for robust statistics at small distances. This approach allows us to compute LSST all-sky efficiencies and determine the total number of expected kilonovae detected during the survey duration (see \autoref{sec:WFD_results}).

\begin{figure*}[!t]
\label{fig:WFD_effs}
\begin{center}
\hspace*{-0.1in}
\scalebox{1.}
{\includegraphics[width=0.6\textwidth]{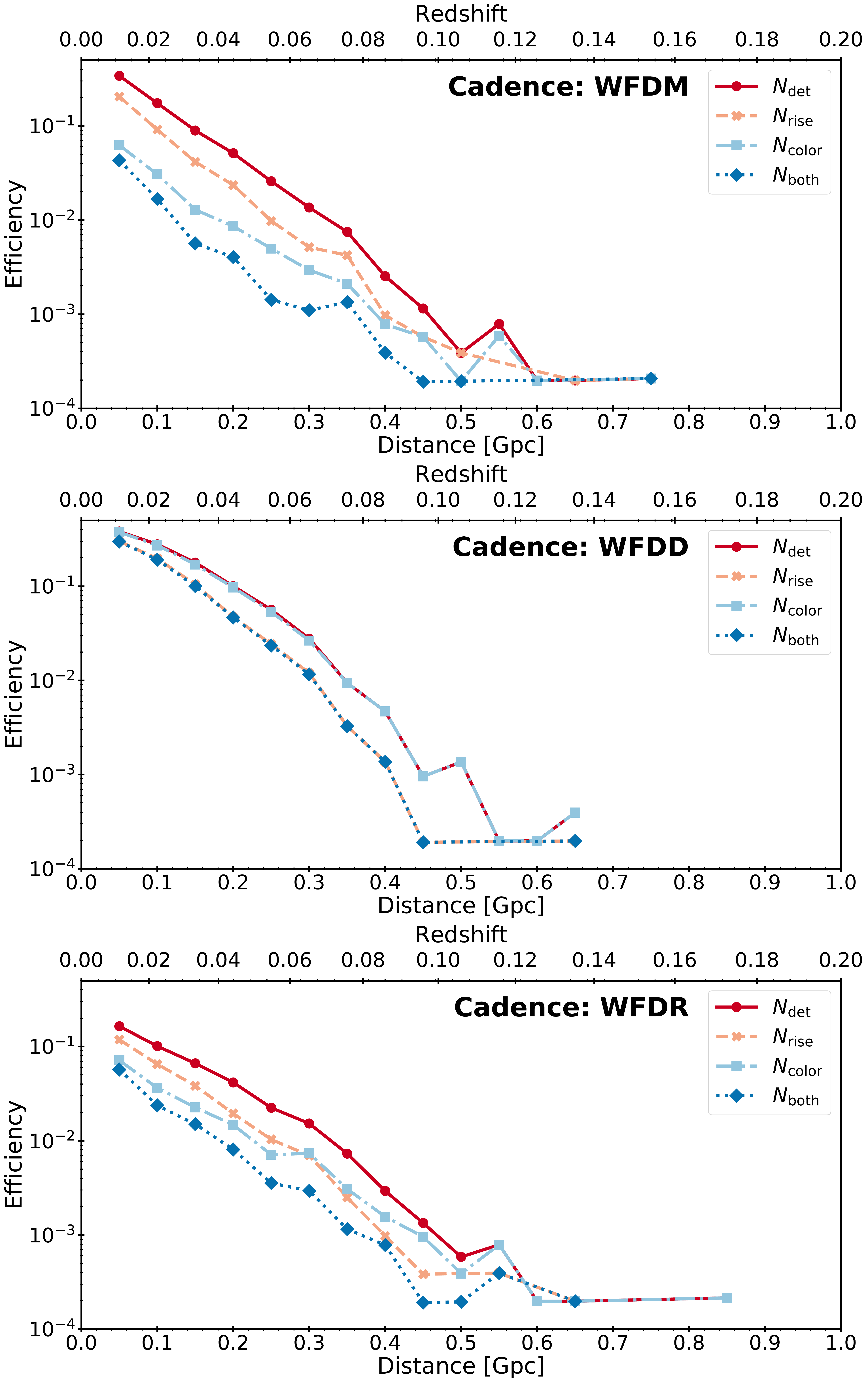}}
\caption{Detection efficiency for kilonovae as a function of distance for each choice of WFD cadence explored in this work. The efficiencies for each criterion: (i) ``detection'' (red), (ii) ``rise'' (orange), (iii) ``color'' (light blue), and (iv) ``both'' (dark blue), are shown individually (see \autoref{sec:WFD_criteria}). We note that for the WFDD cadence the ``detection''/``color'' and ``rise''/``both'' lines overlap as the nightly filter changes allow the light curves to easily satisfy the ``color'' criterion.}
\end{center}
\end{figure*}

\begin{figure*}[!t]
\label{fig:WFD_num}
\begin{center}
\hspace*{-0.1in}
\scalebox{1.}
{\includegraphics[width=0.6\textwidth]{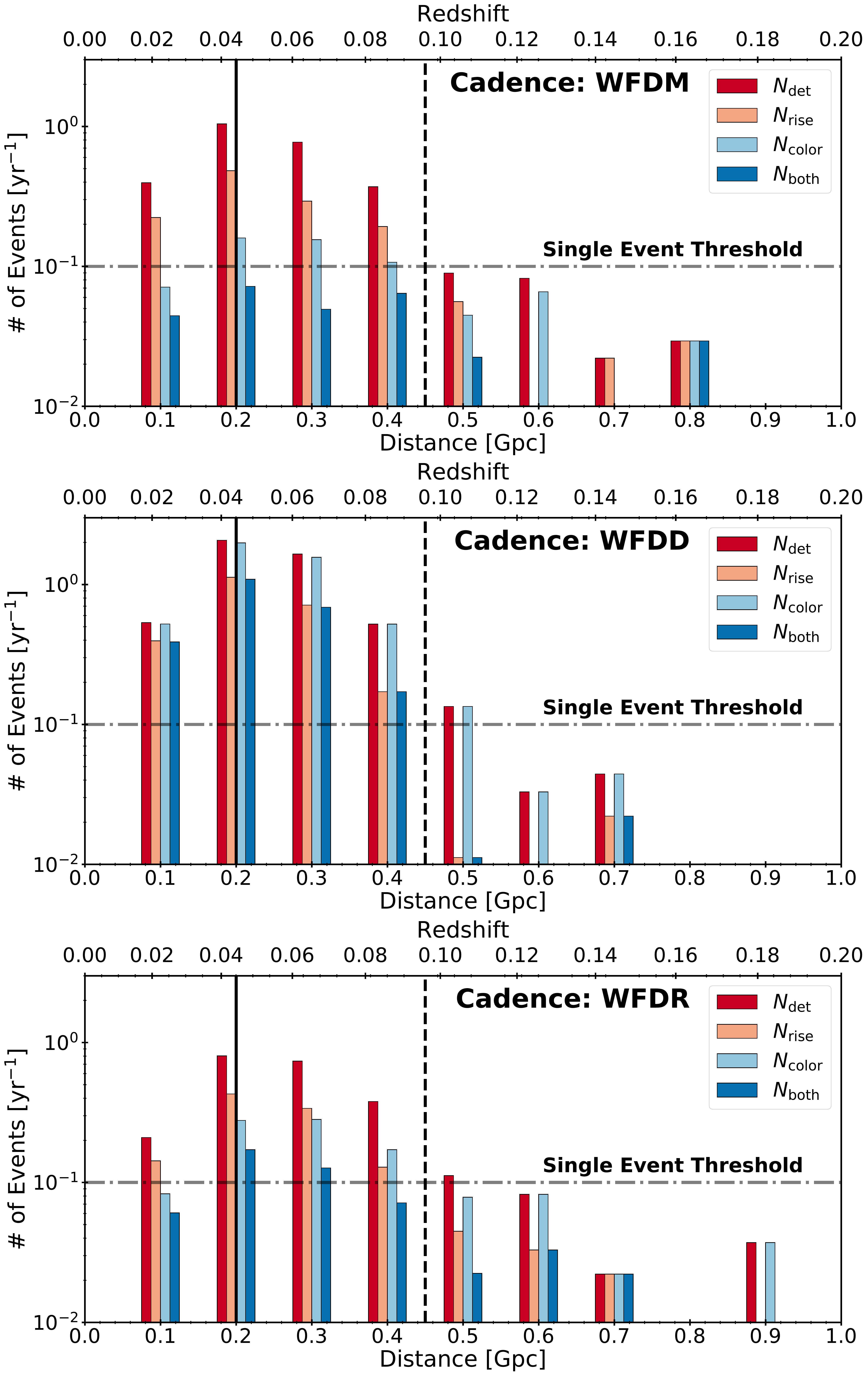}}
\caption{Physical detection rates of kilonovae per year as a function of distance (0.1~Gpc bins) for each choice of WFD cadence explored in this work (see \autoref{sec:WFD_rates}). The detection rates for each criterion: (i) ``detection'' (red), (ii) ``rise'' (orange), (iii) ``color'' (light blue), and (iv) ``both'' (dark blue), are shown individually (see \autoref{sec:WFD_criteria}). The total rate of kilonovae detections is $\apx3-6$ events per year depending on cadence. The vertical black lines indicate distances of 200~Mpc (solid) and 450~Mpc (dashed), the average and maximal ALV BNS detection distances. The horizontal dot-dashed line represents the threshold at which a single event would be detected in the entire ten year survey.}
\end{center}
\end{figure*}

\begin{figure*}[!t]
\label{fig:WFD_det}
\begin{center}
\hspace*{-0.1in}
\scalebox{1.}
{\includegraphics[width=0.75\textwidth]{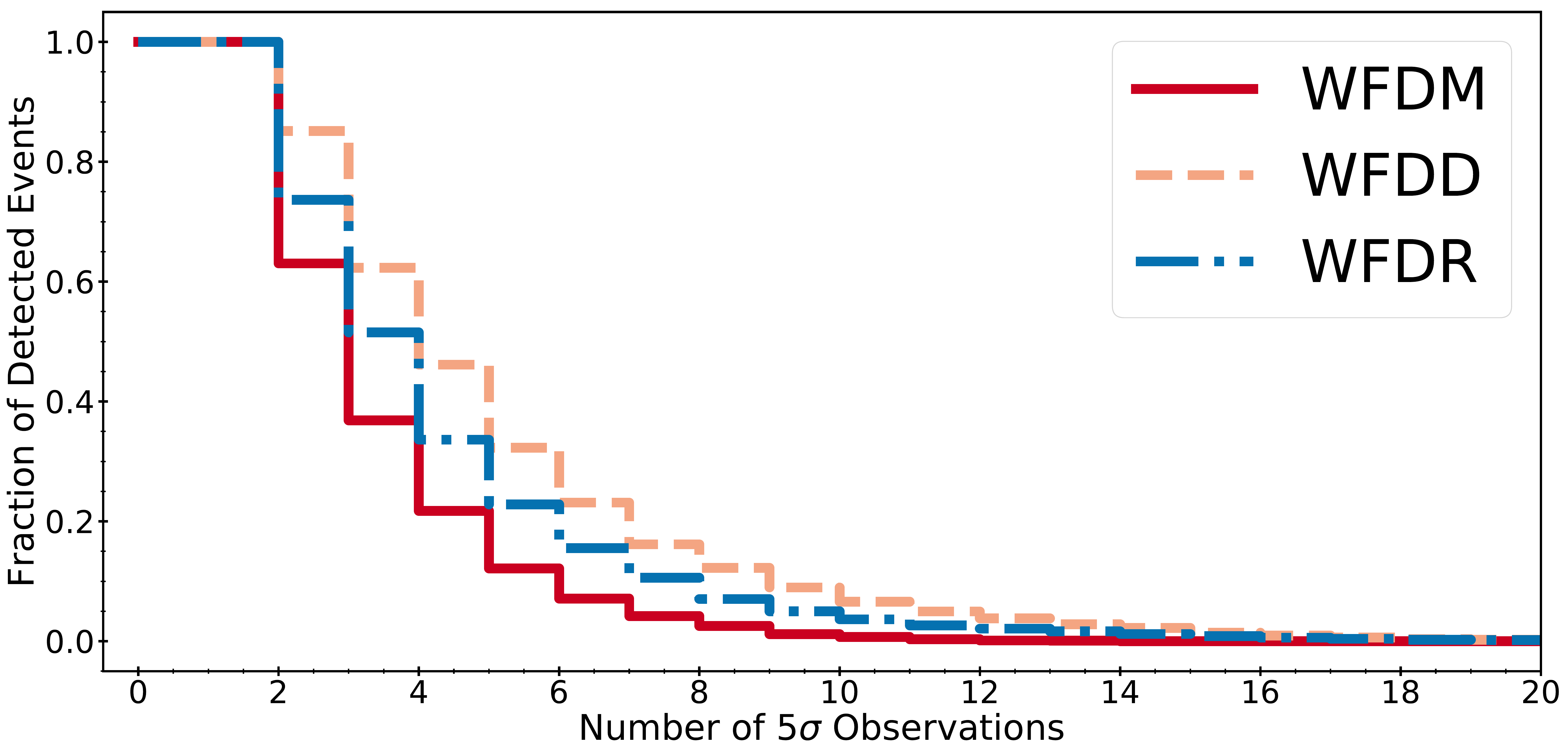}}
\caption{Distribution function showing the fraction of detected kilonovae as a function of number of ${\rm S/N} > 5$ observations for each choice of cadence. Only $\apx10\%$ of sources will have more than $6-8$ data points. This amounts to $\apx3-6$ kilonovae over the ten year survey, depending on the choice of cadence.}
\end{center}
\end{figure*}

\subsection{Target-Of-Opportunity Observations}
\label{sec:obs_too}
We simulate target-of-opportunity observations by constructing custom cadence libraries for use with SNANA. These cadence libraries simulate two sets of $grizY$ observations on the first night after a GW detection, followed by one set per night on the subsequent 6 nights\footnote{We do not simulate $u$-band observations since this filter is less likely to be available.}. The two epochs on the first night are separated by three hours to facilitate studies of the rapid early time evolution. The individual observations in these cadence libraries are built using realistic LSST observing conditions from OpSim, allowing us to explore target-of-opportunity observations across a wide range of realistic observing conditions. Lastly, to assess the effect of integration time on our ToO design, we probe a range of integration times from 30~s (as in the WFD survey) up to 1200~s. 

As described in \autoref{sec:obs_main}, we inject 200,000 model light curves (for each choice of exposure time) into our ToO cadence, with the exception that we use a maximal BNS detection range of $D_{\rm max} \approx 450$~Mpc. The light curves are placed out of phase with the first observation by a uniformly chosen offset time of $3-24$ hr. This is done to represent a realistic range of delays that could affect the start time of the ToO program (e.g., telescope availability, weather, waiting for the target region to become accessible). These considerations allow us to conduct realistic simulations and determine optimal strategies for follow-up (see \autoref{sec:too_results}).

\section{WFD Survey Analysis and Results}
\label{sec:WFD_results}
In this section we investigate two primary questions.  First, the number of kilonovae that will be detected in the WFD survey based on the choice of cadence.  Second, the quality (or information content) of the resulting light curves, and in particular whether they can be used to extract physical information such as mass, velocity, and composition.

\subsection{Detection Criteria}
\label{sec:WFD_criteria}
We first define the following minimal criteria for detection of a kilonova along with key information required for better characterization:

\begin{enumerate}
\item We define a ``detection" as any light curve that has at least 3 observations with ${\rm S/N} > 5$, with at least two of these observations occurring in the same filter. These three observations can be across any combination of times and filters.
\item We define a ``rise" as any light curve that exhibits a $>3\sigma$ increase in flux between two observations in the same filter. This light curve must also satisfy the requirements for a detection. These will be events in which the peak brightness can be roughly estimated.
\item We define a ``color" measurement as any light curve with ${\rm S/N} > 5$ in two independent filters observed within 24 hours of each other. The time constraint is necessary as the colors of kilonovae evolve rapidly. As with the rise criterion, this light curve must also satisfy the requirements for a detection.
\end{enumerate}

We apply these criteria to all 200,000 light curves in our simulated observations. We define the efficiency as the number of kilonovae detected from the total population of injected sources, including sources injected ``off-season" (i.e., at times and sky locations unobservable by LSST). 

We find that the WFDD cadence performs the best with a total efficiency of $\epsilon_{\rm WFDD} = 2.5\%$, while WFDM and WFDR have $\epsilon_{\rm WFDM} = 1.6\%$ and $\epsilon_{\rm WFDR} = 1\%$, respectively. The recovery efficiency has a sharp dependence on the distance of events, and kilonovae beyond $500$~Mpc are generally not detected. The efficiency rises to $\epsilon_{\rm WFDD} = 32\%$, $\epsilon_{\rm WFDM} = 24\%$, and $\epsilon_{\rm WFDR} = 12\%$ for sources within 100~Mpc and $\epsilon_{\rm WFDD} = 11\%$, $\epsilon_{\rm WFDM} = 6.8\%$, and $\epsilon_{\rm WFDR} = 4.3\%$ for sources within 450~Mpc (e.g., the maximal ALV BNS distance). The efficiency curves as a function of distance are shown in \autoref{fig:WFD_effs}.

We also investigate the efficiency at which detected kilonovae satisfy our ``rise" and ``color" criteria. The WFDM cadence results in a factor of two reduction in efficiency for the ``rise" criterion $(\epsilon_{\rm rise, WFDM} = 0.8\%)$ and a further reduction in efficiency for the ``color" criterion $(\epsilon_{\rm color, WFDM} = 0.3\%)$. The efficiency for sources that satisfy both criteria is $\epsilon_{\rm both, WFDM} = 0.2\%$. The WFDR cadence also results in approximately a factor of two reduction in efficiency for all additional criteria with $\epsilon_{\rm rise, WFDR} = 0.6\%$, $\epsilon_{\rm color, WFDR} = 0.4\%$, and $\epsilon_{\rm both, WFDR} = 0.3\%$. Lastly, the WFDD cadence results in a similar drop in efficiency for the ``rise" criterion $(\epsilon_{\rm rise, WFDD} = 1.6\%)$, but the more rapid filter changes result in a negligible reduction in efficiency for the ``color" criterion $(\epsilon_{\rm color, WFDD} = 2.4\%)$. The efficiency for satisfying both conditions is therefore limited by the ``rise" criterion $(\epsilon_{\rm both, WFDD} = 1.6\%)$. These additional efficiencies as a function of distance are shown in \autoref{fig:WFD_effs}.

\begin{figure*}[!t]
\label{fig:WFD_delTHist}
\begin{center}
\hspace*{-0.1in}
\scalebox{1.}
{\includegraphics[width=0.6\textwidth]{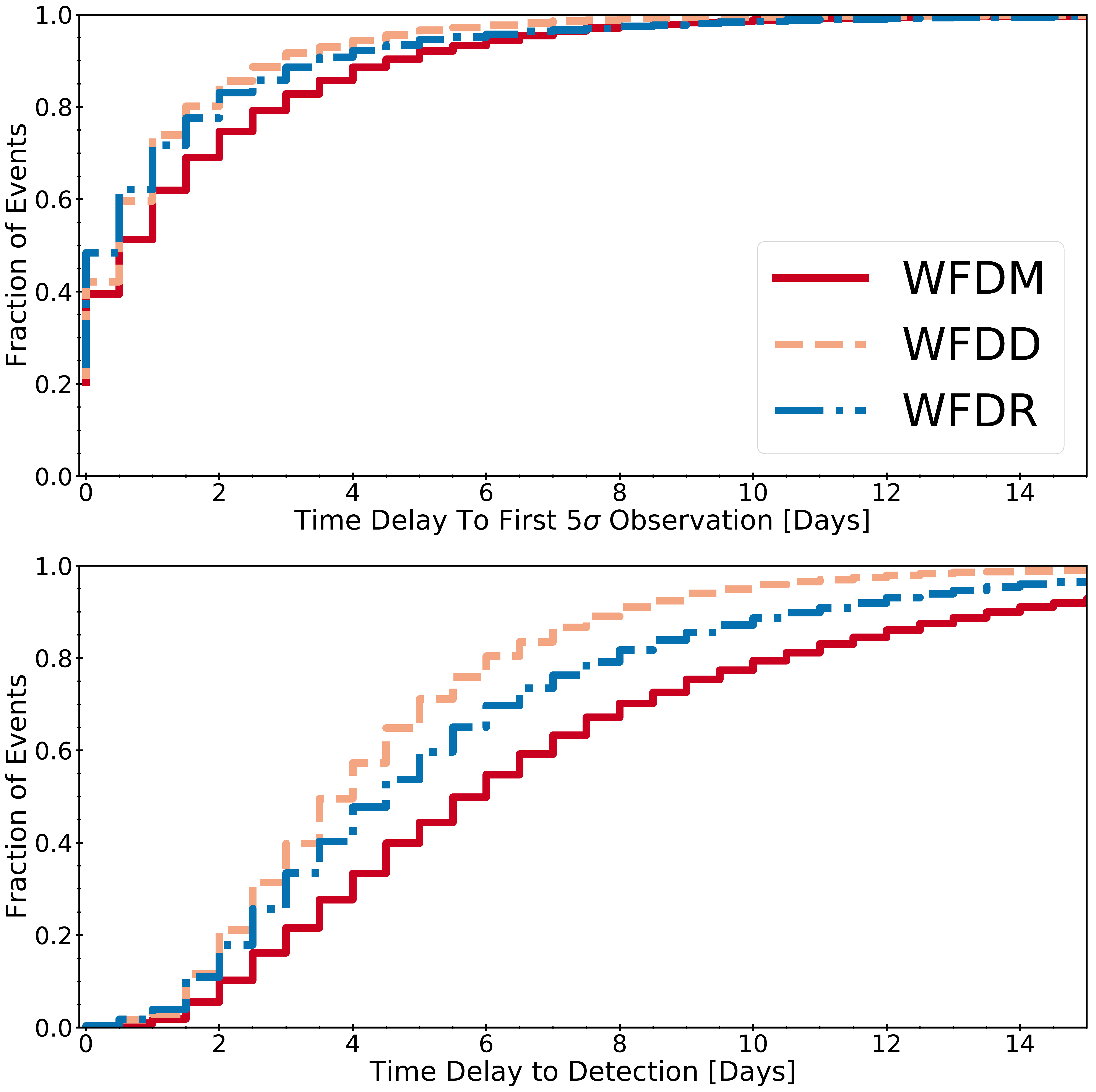}}
\caption{Distribution of the delay time between merger and both the first ${\rm S/N} > 5$ observation (top panel) and meeting our detection criteria (bottom panel), for {\it detected} events. While $40-50\%$ of kilonovae may be observed within one day, the slow WFD cadence means that only $\apx10\%$ will be {\it detected} within two days of merger.}
\end{center}
\end{figure*}

\subsection{WFD Detection Rates}
\label{sec:WFD_rates}
We now compute the total number of kilonovae expected to be detected in the LSST WFD survey. We compute this quantity using the following expression:
\begin{equation}
N_{\rm tot} = \Omega_{\rm LSST} \mathcal{R} \int \epsilon (z) \frac{dV}{dz} dz,
\end{equation}

\noindent where $\mathcal{R}$ is the volumetric rate of kilonovae, $\epsilon (z)$ is the efficiency as a function of redshift, and $dV/dz$ is the differential comoving volume \citep[see e.g.,][]{Hogg1999}. The term $\Omega_{\rm LSST}$ represents the total LSST survey area ($\Omega_{\rm LSST} \approx 18,000$ deg$^2$). We assume the rate of kilonovae is the rate of BNS mergers derived by LIGO during the second observing run \citep[$\mathcal{R} \apx1500$~Gpc$^{-3}$~yr$^{-1}$,][]{LIGOGW170817}. We assume that $\mathcal{R}$ does not evolve over the relevant small redshift range. We additionally note that while the merger rate still has large uncertainties, our efficiency as a function of distance is robust.

We perform the above calculation for each choice of cadence. We find that the WFDD cadence performs the best with an expected rate of $N_{\rm tot, WFDD} \approx 6.6$ detections per year (i.e., sources that satisfy criterion 1). The WFDM and WFDR cadences lead to $N_{\rm tot, WFDM} \approx 3.7$ and $N_{\rm tot, WFDR} \approx 3.0$ detections per year, respectively. In all three cases these numbers are lower than those found by \citet{Scolnic+18}, but this is readily attributed to the stricter detection criteria utilized here (which we stress are still rather forgiving).

We also compute detection rates for each choice of cadence using the efficiencies computed for the ``rise'' and ``color'' criteria. We find that the WFDD cadence continues to perform the best with $N_{\rm rise, WFDD} \approx 3.4$, $N_{\rm color, WFDD} \approx 6.4$, and $N_{\rm both, WFDD} \approx 3.3$ detections per year. The detection rates for the WFDR cadence are $N_{\rm rise, WFDR} \approx 1.5$, $N_{\rm color, WFDR} \approx 1.3$, and $N_{\rm both, WFDR} \approx 0.6$ detections per year. The WFDM cadence performs the worst with $N_{\rm rise, WFDM} \approx 1.7$, $N_{\rm color, WFDM} \approx 0.8$, and $N_{\rm both, WFDM} \approx 0.4$ detections per year. This indicates that the number of detected kilonovae with useful scientific information (i.e. rise, peak, and color) will be low, regardless of cadence.

Lastly, we compute the detection rate for events found within the ALV detection volume. Here we just focus on events that satisfy the first detection criterion within 200~Mpc and 450~Mpc, the average and maximal BNS detection distances. Considering sources within 200~Mpc, we find $N_{\rm 200~Mpc, WFDD} \approx 3.4$, $N_{\rm 200~Mpc, WFDM} \approx 1.9$, and $N_{\rm 200~Mpc, WFDR} \approx 1.3$ detections per year. For sources within 450~Mpc we find $N_{\rm 450~Mpc, WFDD} \approx 6.5$, $N_{\rm 450~Mpc, WFDM} \approx 3.5$, and $N_{\rm 450~Mpc, WFDR} \approx 2.9$ detections per year. This indicates that the bulk of kilonovae expected to be detected by the WFD survey will be within the ALV detection volume.

\begin{figure*}[!t]
\label{fig:WFDM_fits}
\begin{center}
\scalebox{1.}
{\includegraphics[width=0.65\textwidth]{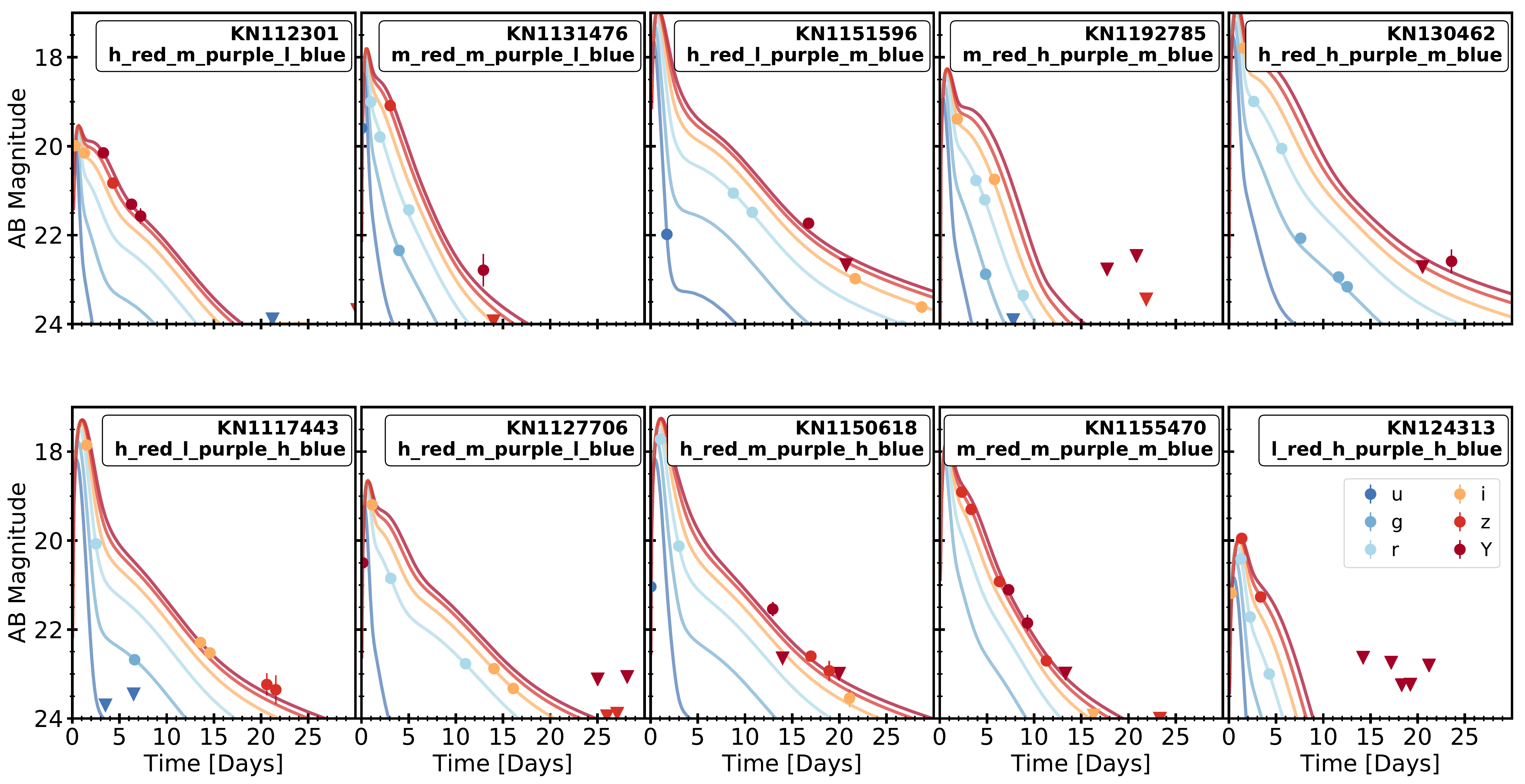}}
{\includegraphics[width=0.65\textwidth]{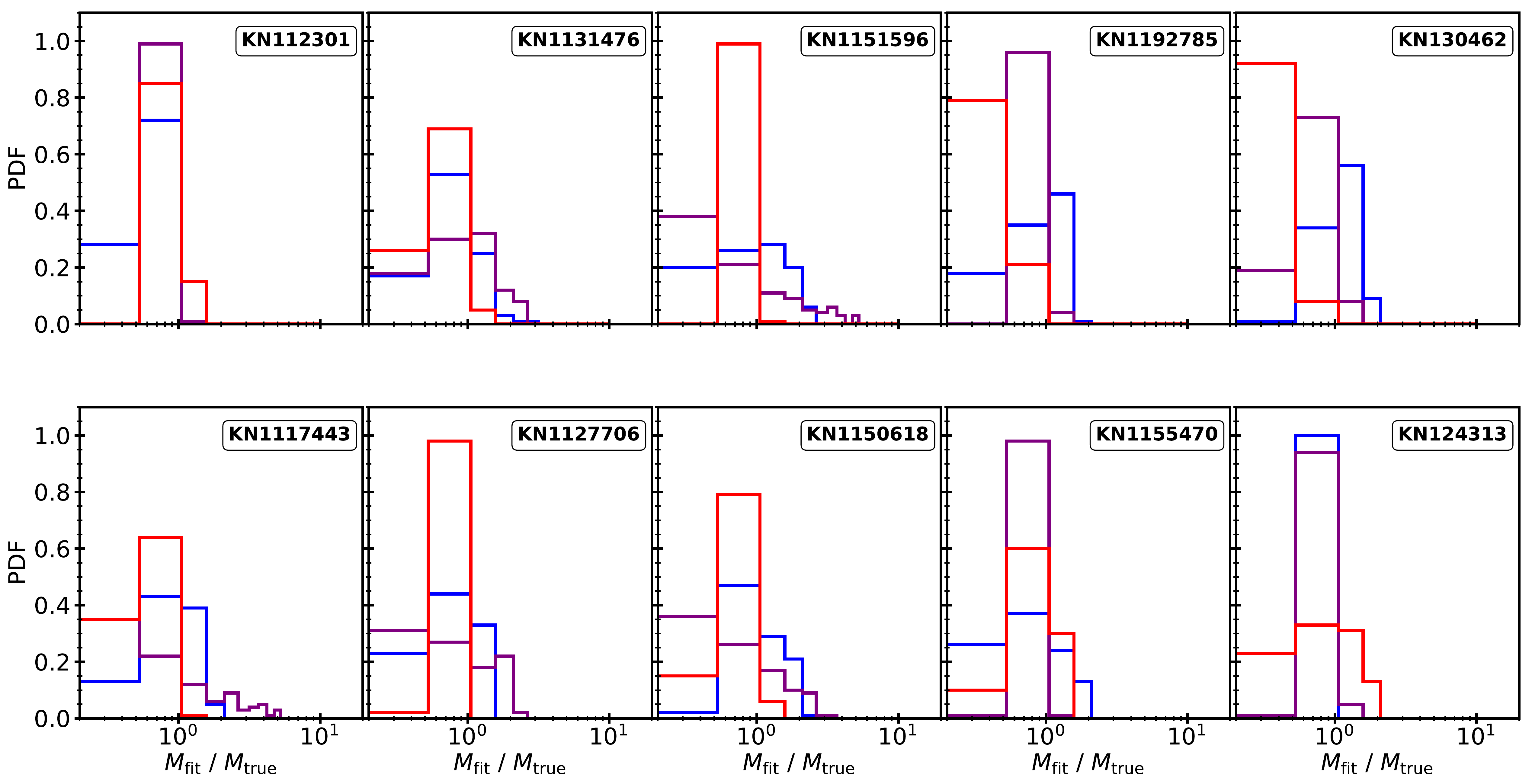}}
{\includegraphics[width=0.65\textwidth]{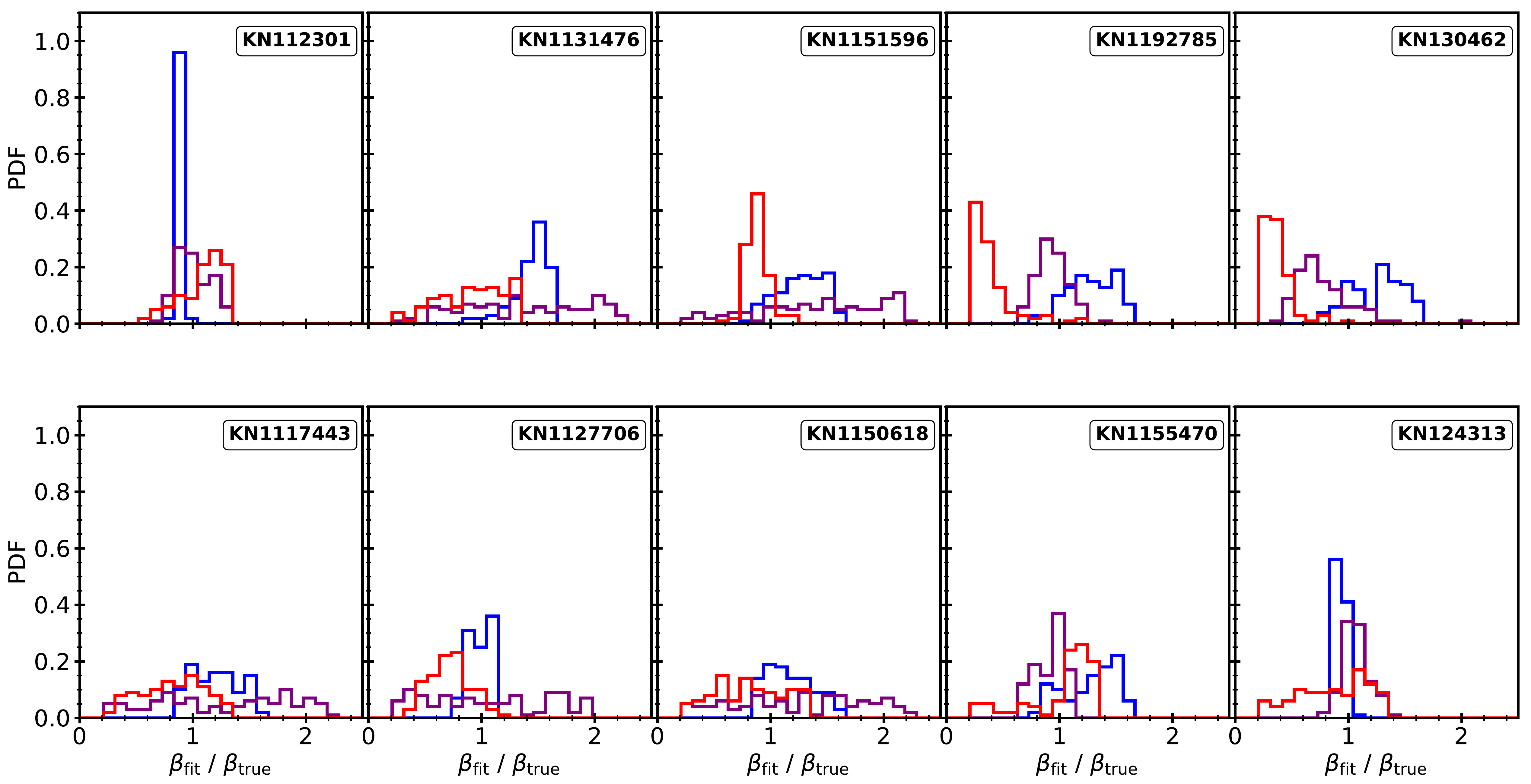}}
\caption{{\it Top Two Rows:} Simulated observations and model light curves for kilonovae in the WFDM cadence. These light curves represent the top 10\% of all detected events in terms of the number of data points (representing $3-6$ events in the entire ten year survey). {\it Middle Two Rows:} Posteriors for ejecta mass as a function of component, relative to the injected value. {\it Bottom Two Rows:} Posteriors for ejecta velocity as a function of component, relative to the injected value.}
\end{center}
\end{figure*}

\subsection{Kilonova Data Quality}
\label{sec:WFD_data}
We next investigate the data quality (beyond the simple "detection" criterion) of the detected kilonovae. First, we investigate the distribution of events as a function of the number of observations with ${\rm S/N} > 5$, independent of filter.  Second, we investigate the distribution of time delay between when the light curve is injected into the WFD survey ($t_{\rm injection}$) and the first ${\rm S/N} > 5$ detection, in any filter, for a given event. We additionally compute the delay time between $t_{\rm injection}$ and when an event satisfies our detection criterion.

In \autoref{fig:WFD_det} we show the cumulative distribution of detected events for each cadence as a function of the number of ${\rm S/N} > 5$ observations. We find, across all choices of cadence, that 80\% of events will have fewer than five to seven ${\rm S/N} > 5$ observations across all filters. Furthermore, only the WFDD and WFDR cadences produce any light curves with more than 20 ${\rm S/N} > 5$ observations. However, these represent only the top $\approx0.5\%$ of events. Given the low detection rates inferred in \autoref{sec:WFD_rates}, the likelihood of such events being found in the WFD survey is low. Therefore, the majority of detected kilonovae will have light curves that are too sparsely sampled to extract useful information about the event (see \autoref{sec:WFD_modeling}). 

In \autoref{fig:WFD_delTHist} we show the cumulative distribution function as a function of both the time to reach a "detection" (three ${\rm S/N} > 5$ observations) and the time delay between $t_{\rm injection}$ and the first ${\rm S/N} > 5$ observation. We find that 50\% of sources will be observed within 1 day of $t_{\rm injection}$ for the WFDM and WFDD cadences and within 12 hours for the WFDR cadence. All three cadences will allow LSST to observe 90\% of events within $3-4$. However, these events will not satisfy our detection criterion until much later. Independent of cadence, only $\approx50\%$ of kilonovae satisfy our detection criterion after $3-5$ days, while 10\% of kilonovae do not satisfy our detection criterion until $>2$ weeks. This delay will reduce the effectiveness of triggering follow-up observations (e.g., spectroscopic and multi-wavelength observations) for detected events. These additional observations are crucial for building a complete understanding of kilonovae. 

\begin{figure*}[!t]
\label{fig:WFDD_fits}
\begin{center}
\scalebox{1.}
{\includegraphics[width=0.65\textwidth]{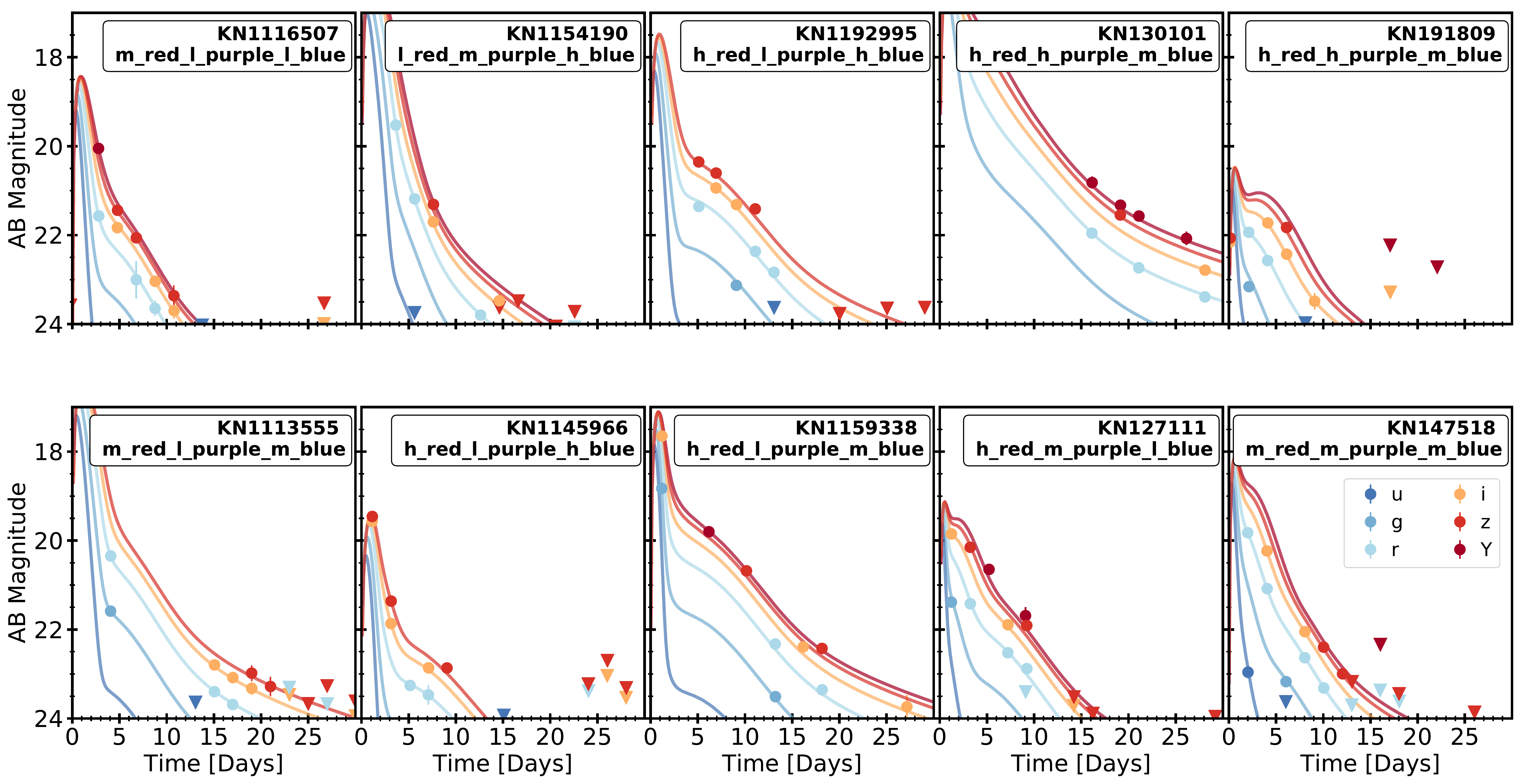}}
{\includegraphics[width=0.65\textwidth]{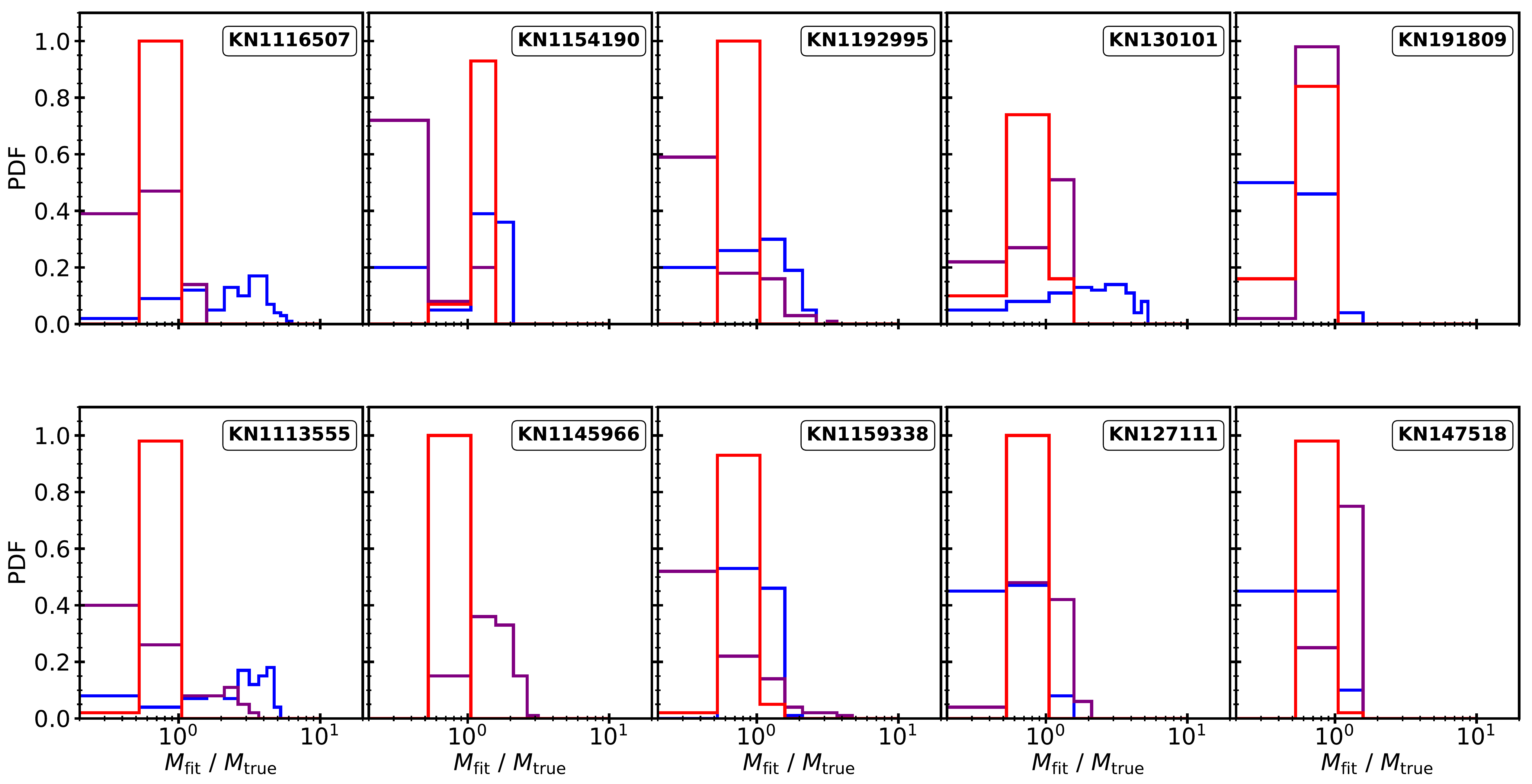}}
{\includegraphics[width=0.65\textwidth]{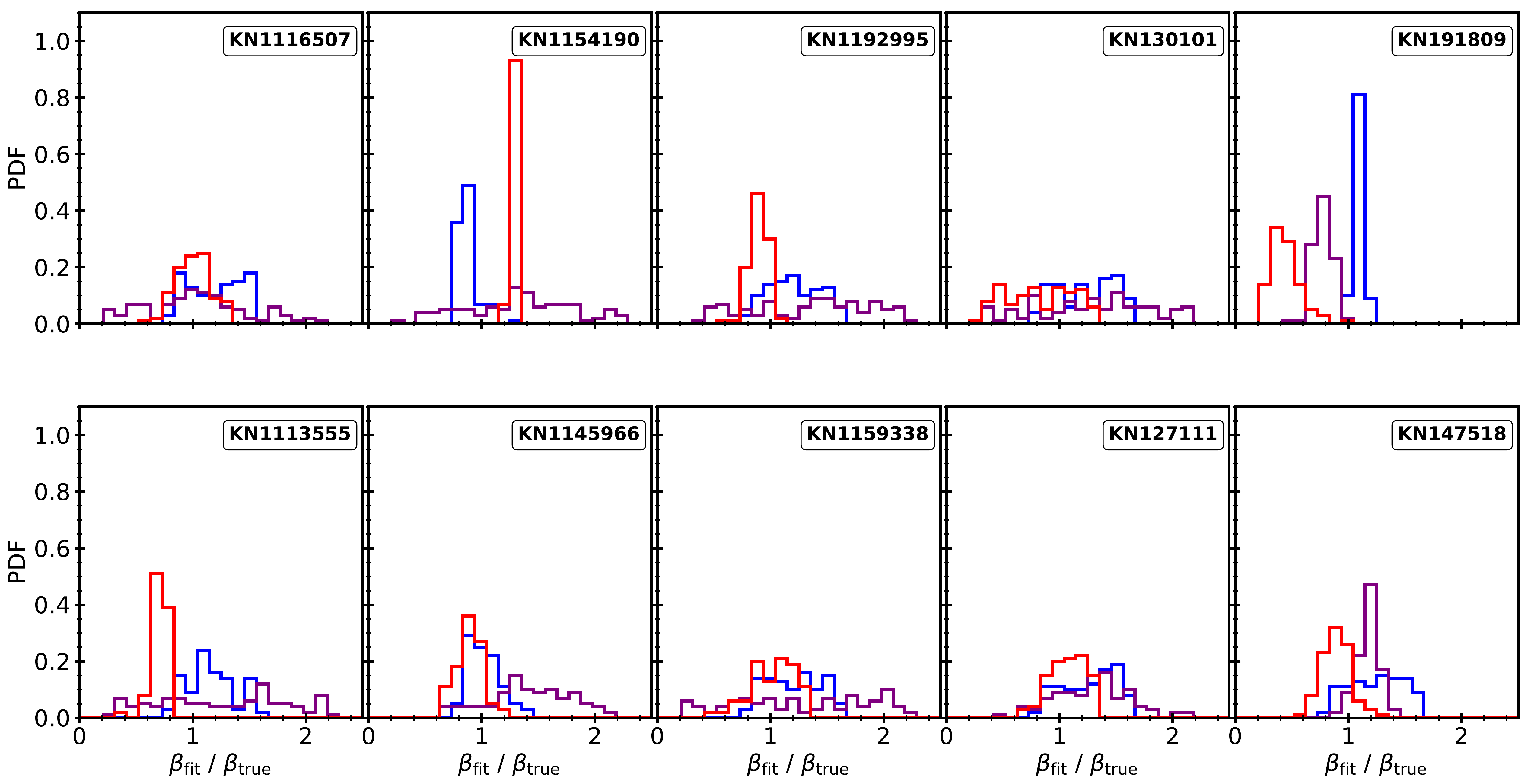}}
\caption{As \autoref{fig:WFDM_fits}, but for the WFDD cadence.}
\end{center}
\end{figure*}

\begin{figure*}[!t]
\label{fig:WFDR_fits}
\begin{center}
\scalebox{1.}
{\includegraphics[width=0.65\textwidth]{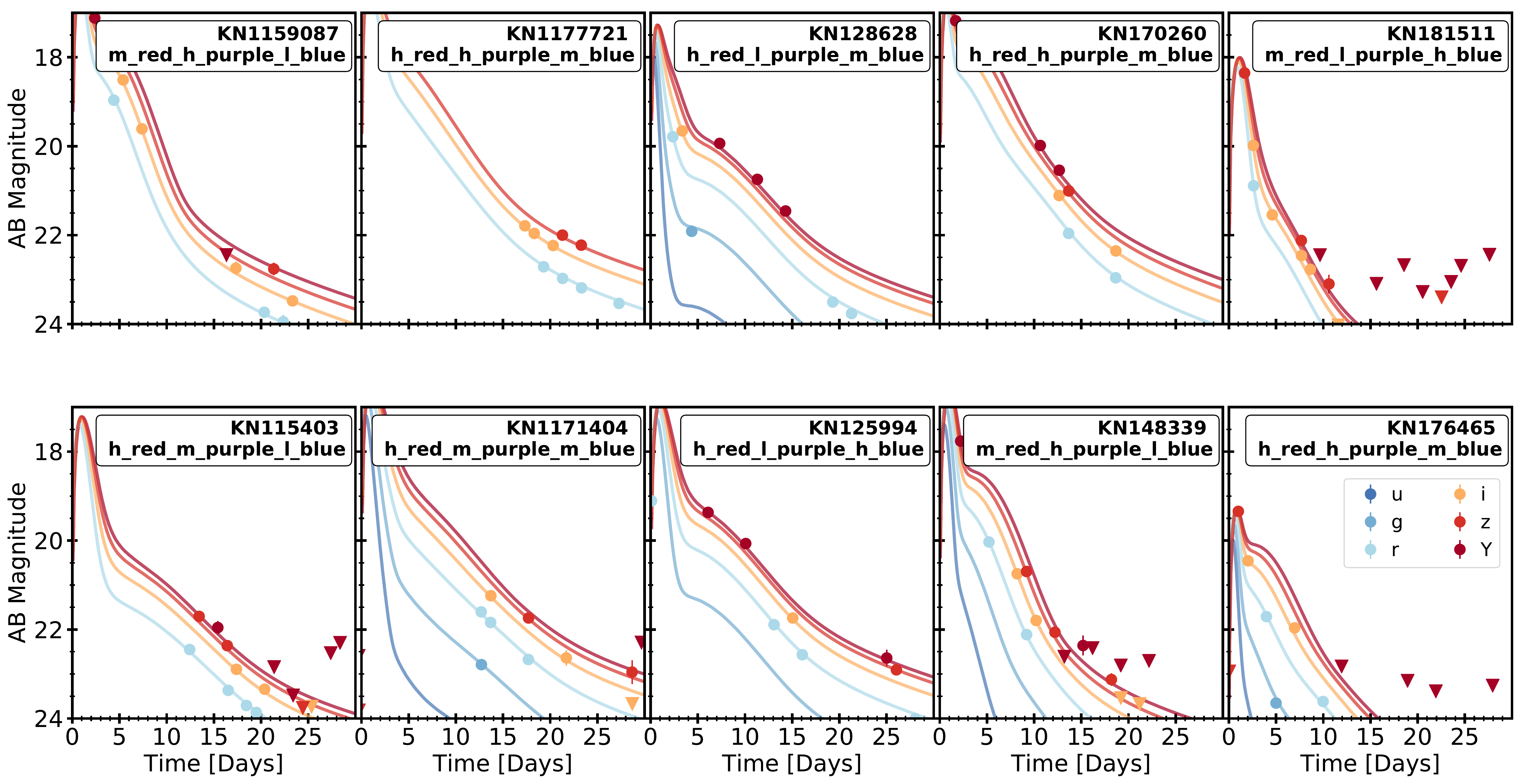}}
{\includegraphics[width=0.65\textwidth]{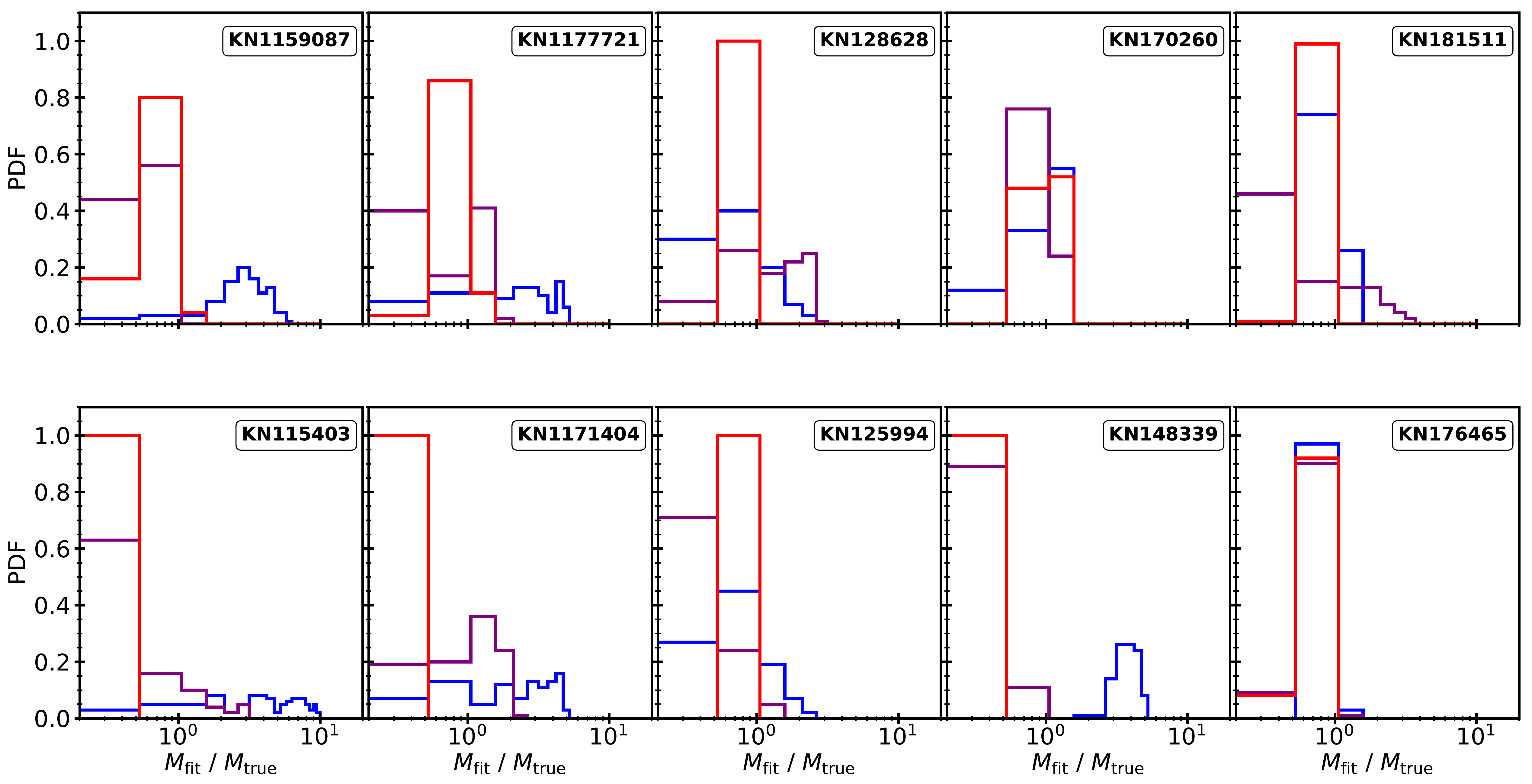}}
{\includegraphics[width=0.65\textwidth]{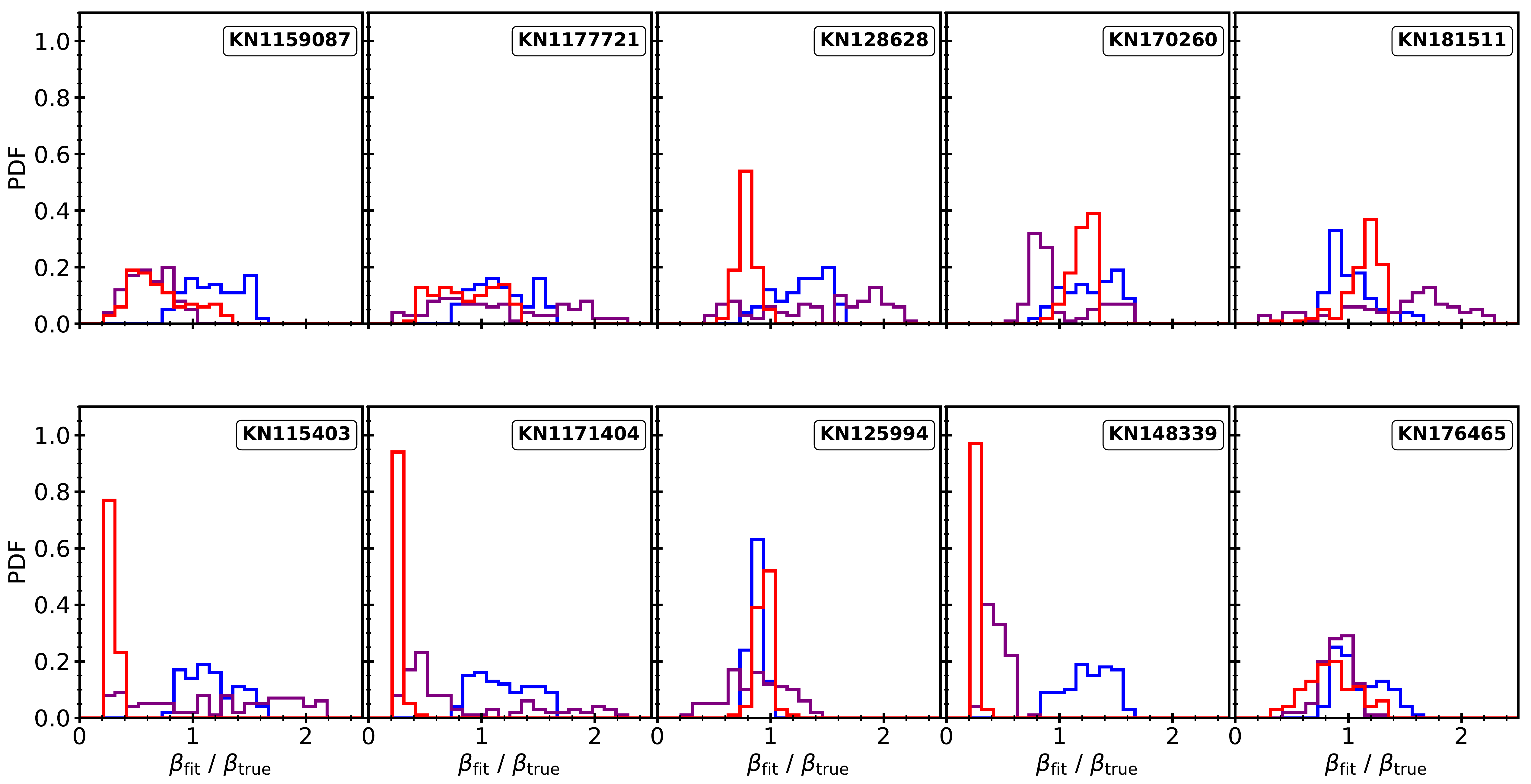}}
\caption{As \autoref{fig:WFDM_fits}, but for the WFDR cadence.}
\end{center}
\end{figure*}

\begin{figure*}[!t]
\label{fig:mass_fitted}
\begin{center}
\hspace*{-0.1in}
\scalebox{1.}
{\includegraphics[width=0.6\textwidth]{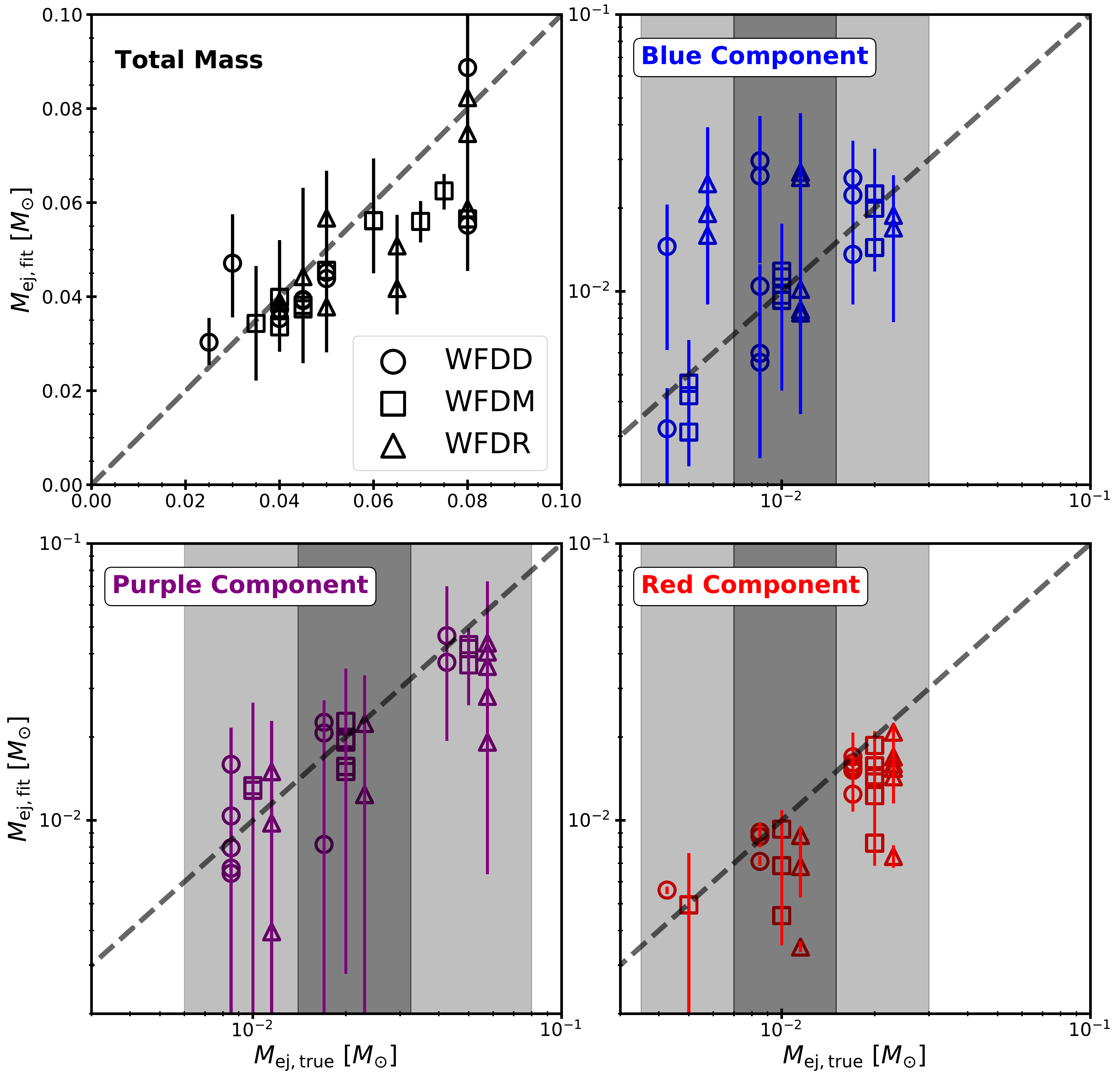}}
\caption{Comparison of the fitted versus injected ejecta masses for each of the three kilonova components and each cadence. A slight horizontal shift is applied to the WFDD and WFDR cadences for visibility. The gray shaded regions indicate each choice of component mass as outlined in \autoref{sec:models}. The total ejecta mass is shown in the top left panel.}
\end{center}
\end{figure*}

\begin{figure*}[!t]
\label{fig:vel_fitted}
\begin{center}
\hspace*{-0.1in}
\scalebox{1.}
{\includegraphics[width=0.6\textwidth]{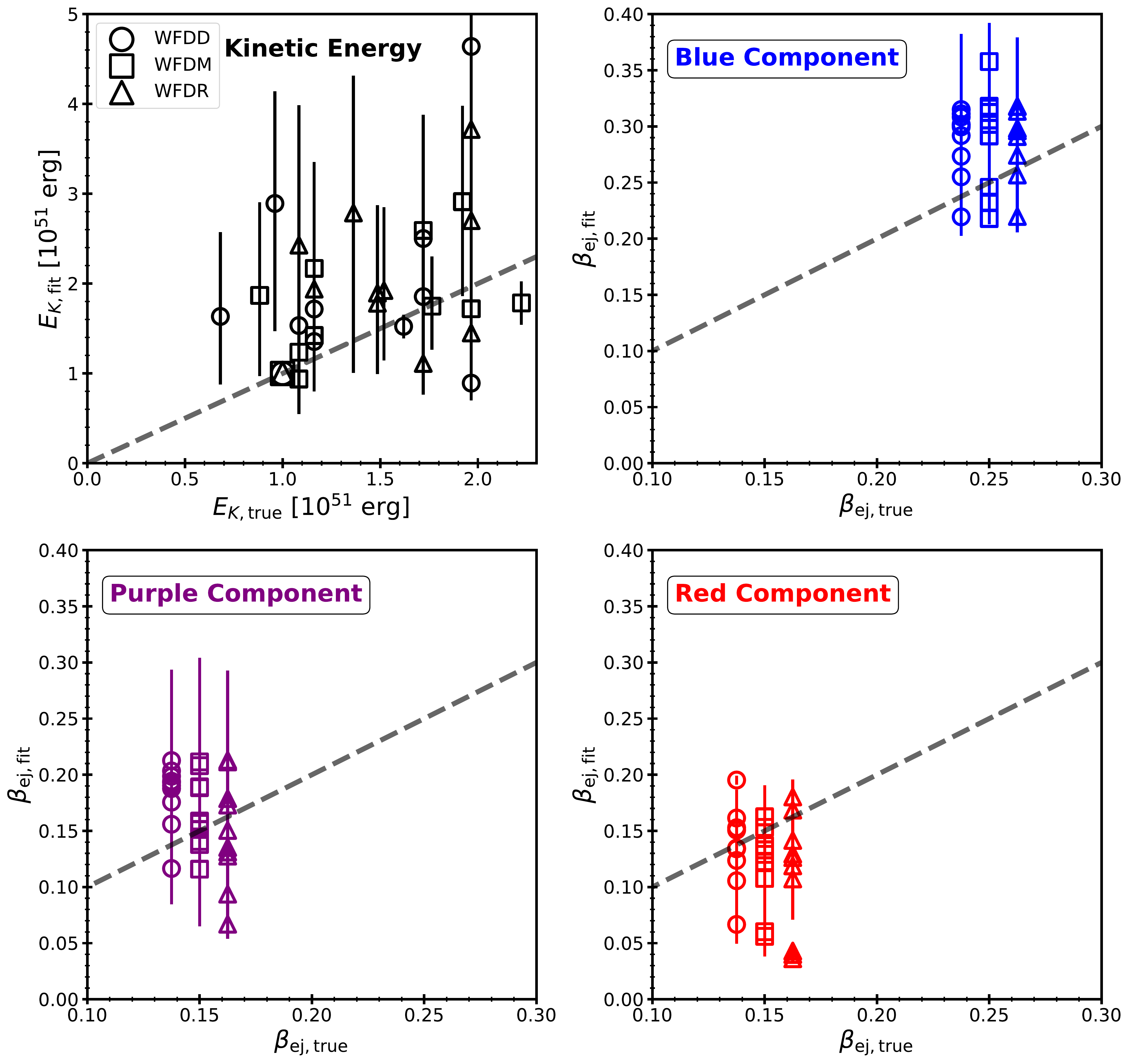}}
\caption{Same as \autoref{fig:mass_fitted}, but for the ejecta velocity of each component. The kinetic energy of the ejecta is shown in the top left panel.}
\end{center}
\end{figure*}

\subsection{Modeling the Light Curves of Recovered Kilonovae}
\label{sec:WFD_modeling}
Lastly, we investigate our ability to extract key model parameters for detected kilonovae. As stated in \autoref{sec:WFD_data}, accurate modeling of the light curves is essential to maximize the science return from a given event. Here, we are primarily focused on the accuracy of the recovered ejecta mass ($\Mej$) and ejecta velocity ($\vej$) to facilitate robust comparison to simulations. Specifically, accurate determination of $\Mej$ provides insight into mergers involving neutron stars as primary sites of cosmic $r$-process production, the physical origin of ejecta (e.g., tidal vs. winds), and information about the binary components and remnant compact object. Likewise, $\vej$ is a crucial diagnostic for determining physical mechanisms for ejecta components.

To provide a view of the best case scenario, we randomly select 10 light curves (from each cadence) in the 90th percentile of events based on the number of ${\rm S/N} >= 5$ observations. Depending on the individual efficiency of each cadence (see \autoref{sec:WFD_rates}), these light curves are a representative sample of the best $3-6$ events from the entire 10-year WFD survey. We refit each light curve with \mosfit, applying the same three-component model that was used to generate the light curves. This model has 7 free parameters (ejecta mass and velocity per component, and a single scatter term). We stress that since these represent the highest quality light curves, all of the events actually occur within the ALV BNS detection volume. Therefore we assume that the distance and merger time are known and can be fixed during fitting. Additionally, the temperature floor parameter is fixed at the injected value to reduce the number of free parameters, as it is most relevant at later times. We do not make any additional assumptions about the intrinsic nature of the source or ejecta properties and do not alter the priors based on our knowledge of the injected model. We run each model to convergence as described in \citet{Villar+17b}. The data and best fit light curves are shown in the top two rows of \autoref{fig:WFDM_fits} (WFDM Cadence), \autoref{fig:WFDD_fits} (WFDD Cadence), and \autoref{fig:WFDR_fits} (WFDR Cadence). 

The fitted mass ($M_{\rm fit}$) posteriors scaled to the injected ejecta mass ($M_{\rm true}$) are shown in the middle two rows of \autoref{fig:WFDM_fits} (WFDM Cadence), \autoref{fig:WFDD_fits} (WFDD Cadence), and \autoref{fig:WFDR_fits} (WFDR Cadence). We find that, in general, the recovered masses differ from the injected values by a factor of $2-3$, but with large uncertainties. For the worst events, the fitted mass can be off by a factor of $6-8$. This is also shown in \autoref{fig:mass_fitted}, in which we plot the injected versus fitted ejecta masses for each component, as well as the total ejecta mass. We find that the total ejecta mass for the system is generally underestimated by less than a factor of 2. Lastly, we note that the more accurately determined masses come from light curves with observations before or around the peak, highlighting the importance of early detection. 

The fitted velocity ($v_{\rm fit}$) posteriors scaled to the injected velocity ($v_{\rm true}$) are shown in the bottom two rows of \autoref{fig:WFDM_fits} (WFDM Cadence), \autoref{fig:WFDD_fits} (WFDD Cadence), and \autoref{fig:WFDR_fits} (WFDR Cadence). We find that in general, the ejecta velocity is similarly constrained to within a factor of $2-3$, but again with large uncertainties. This is shown on a per component basis in \autoref{fig:vel_fitted}. We find that the velocity of the faster ``blue'' component is typically over-estimated, while the velocity of the slower ``red'' component is systematically underestimated. This results in the total kinetic energy of the ejecta being overestimated, as seen in the top left panel of \autoref{fig:vel_fitted}. Unlike with the connection between observations around peak and accurate determination of the ejecta mass, there is no obvious link between the qualitative properties of the light curve and the accuracy of the fitted velocities.

\subsection{Final Considerations}
\label{sec:WFD_considerations}
The key point emerging from this study is that despite the large \'{e}tendue of LSST, the WFD survey will be relatively inefficient at finding kilonova, even with a loose definition of detection as 3 data points. This is true independent of the survey cadence for the realistic choices tested in this work. More importantly, we have shown that only $\apx10\%$ of  detected kilonovae will have sufficient light curve data for more detailed studies. This is approximately $3-6$ events over the entire survey duration all of which are sufficiently nearby to be detected by ALV anyway. However, even in these most optimistic cases it is not always the case that ejecta parameters can be accurately characterized to better than a factor of $2-3$. This suggests that the science returns from the meager LSST kilonova sample will be minimal.

The low efficiency of the main survey also suggests that the likelihood of a serendipitous joint detection between ALV and LSST is vanishingly small. While LSST may observe a reasonable fraction $(40-50\%)$ of kilonovae at early-times, the slow cadence results in only a small fraction of events being detected ($\apx 10\%$ within two days of merger). These delays will severely limit the ability to perform multi-wavelength follow-up observations, which were found to be crucial for maximizing our understanding of GW170817.

\section{LSST Target of Opportunity Observations of GW Triggers}
\label{sec:too_results}
Given the shortcomings for kilonova identification and study in the WFD survey, we now turn to an investigation of ToO observations of GW-triggered events. In this section, we outline several key considerations required for the development of a successful ToO program. The principle guiding these considerations is the exploration of a minimal set of observations that can rapidly identify kilonovae, followed by observations with other facilities for further observation. These considerations include: exposure design (integration time, choice of filter, etc.), timing (cadence, promptness of response times), and source identification (detection criteria, contaminant rejection). Of course, more time can be allocated in favorable or unique circumstances (see \autoref{sec:too_early}), but the goal here is to minimize the impact on the baseline LSST operations. These suggestions represent the crucial first steps towards the development of an effective and efficient GW ToO program with LSST. 

\begin{figure*}[!t]
\label{fig:too_effs}
\begin{center}
\hspace*{-0.1in}
\scalebox{1.}
{\includegraphics[width=0.75\textwidth]{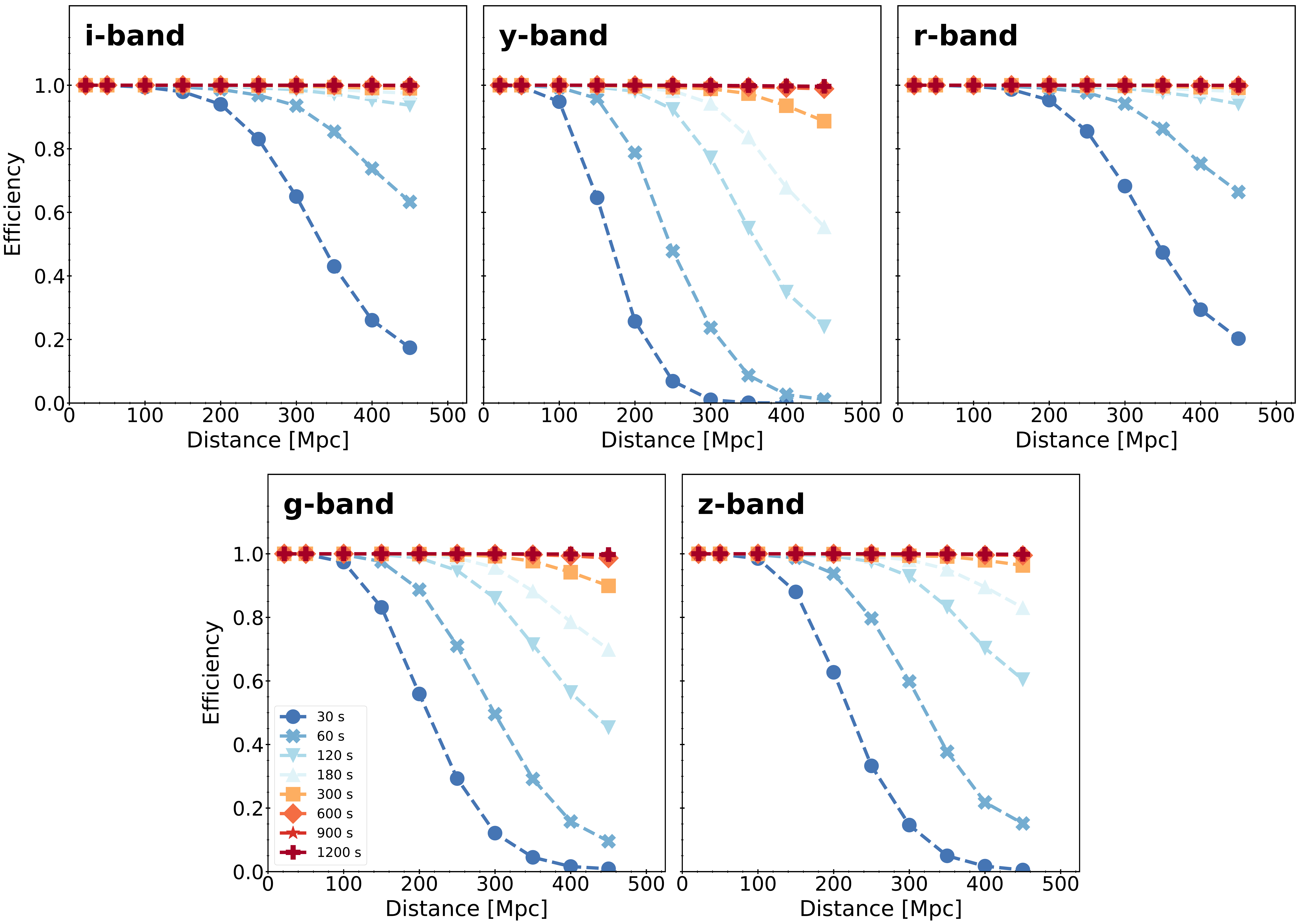}}
\caption{Detection efficiency for kilonovae in our ToO simulation as a function of filter, integration time, and distance.}
\end{center}
\end{figure*}

\vspace{10pt}
\subsection{Requirements for Detection}
\label{sec:too_criteria}
We first address the issue of selecting ToO specific detection criteria. In particular, we investigate the ability of LSST to rapidly detect kilonovae. The light curve evolution in both timescale and luminosity is a strong function of filter. Therefore, we investigate the exposure time necessary to obtain robust detections in each filter. We accomplish this by running our ToO simulations (\autoref{sec:obs_too}) using a grid of integration times spanning from $30-1200$~s. We then investigate what fraction of kilonovae are detected at ${\rm S/N} \ge 5$, at least three times, in each filter \footnote{Here we do not consider $u$-band as it is less likely to be available for ToO observations and is overall less efficient.}. 

The efficiency curves, per filter, as a function of distance are shown in \autoref{fig:too_effs}. We find that integration times of $120$~s are sufficient to obtain robust early-time detections of kilonova to 200~Mpc (the average ALV BNS detection distance) across all filters. However, for redder filters (e.g., $i$ and $z$), integrations times as short as 60 s are sufficient to achieve a $90\%$ detection efficiency. Achieving a similar detection efficiency for more distant sources (e.g., favorably aligned BNS mergers or NS-BH mergers), will require integration times of $180-300$~s depending on the choice of filter. Lastly, we note that since our simulations use realistic outputs from OpSim, these efficiencies include losses due to weather and other environmental factors (e.g., moon brightness). Therefore, while the integration time could in practice be adjusted depending on the exact distance and observing conditions, we will use a fiducial integration time of $180$~s in this discussion. 

\subsection{Choice of Filters}
\label{sec:too_filters}
We next discuss the choice of filters required to make rapid and robust detections of kilonovae in LSST observations. In this context of prompt discovery, it is not necessary nor efficient to conduct observations in all available bands. Rather, we argue that observations in $g$ and $z$-bands are sufficient to capture the range of kilonovae behavior required for rapid identification. Broadly speaking, the red colors of kilonovae are a strong discriminator against other optical transients \citep[see e.g.,][]{CowpBerger15,Doctor+17}. For example GW170817, despite the unexpectedly bright blue component, was redder than supernovae even at early times \citep[e.g., $\lesssim12$~hours, see][]{Cowp+17}.

More specifically, the choice of $g+z$ observations will allow observations to probe the near extremes of the optical SED, without the possible loss of throughput in $u$ and $Y$-bands. This will capture unique light curve evolution across a diverse range of possible kilonova models. For example, the models used in this work exhibit $g-z$ colors that redden from $g-z \approx -0.5$~mag at $\apx 3$~hours to $g-z \approx 0.2$~mag at $\apx12$~hours and further redden to $g-z \approx 1$~mag in $\apx1$~day. In the absence of blue emission, the source will promptly exhibit rapid color evolution and likely non-detections in $g$-band.

\subsection{Impact of Rapid Follow-Up on Cadence}
\label{sec:too_early}
A key strength of LSST targeted follow-up will be its ability to place deep constraints on the early behavior of kilonovae in circumstances where the sky position and time are favorable for prompt follow-up. If the target field can be observed within minutes of merger, then a monitoring program should be initiated in the $g,z$ filters with a cadence that is a doubling of the response time (when applicable). For example, if observations of the field commence at ten minutes post-merger, the program cadence would have observations at +10, +20, +40, +80, etc., until identification (ideally no more than five epochs). These observations would capture the kilonova as a rapidly rising source with an evolution that is much more rapid than typical supernovae. Such observations will be essential for understanding the early behavior of kilonovae and unveiling new emission mechanisms (e.g., shock cooling \citealt{PiroKollmeier18} or a neutron ``precursor" \citealt{Metzger+15}).

In scenarios where rapid response is not possible, a successful ToO strategy should still strive to obtain two or more sets of observations per night, separated by $\approx3$ hours. The rapidly changing nature of kilonovae light curves means that observations taken even 24 hours apart may capture very different stages of evolution. This was seen in GW170817 where $g$-band observations revealed a much hotter and bluer transient than similar observations taken just a few hours later \citep[see e.g.,][]{Cowp+17,Drout+17}. Multiple observations on the same night will also allow more rapid identification of kilonovae candidates, allowing more effective follow-up with complimentary facilities and potentially reducing the overall impact on LSST operations.

Ultimately, the choice of ToO cadence will be largely determined by the response time. In the case of possible rapid response (e.g., within minutes), an intra-night doubling cadence with about five epochs will provide unprecedented insight into the early behavior of kilonovae, while at the same time minimizing the impact on LSST operations. This will allow the counterpart to be identified and monitored while narrow-field telescopes take over the longer-term monitoring.  On the other hand, if only a delayed response is possible (due to observing constraints) then two epochs on the first night (separated by a few hours), followed by a possible pair of epochs on the second night will ensure discovery of the counterpart (at which point it can be followed up by narrow-field telescopes).  Thus, in either regime we anticipate that at most $4-5$ epochs of observations will be required, minimizing the impact on LSST operations.

\subsection{Contamination Rejection}
\label{sec:too_contamination}
An important factor impacting our ability to rapidly detect and identify kilonovae is contamination from unrelated background and foreground sources. While the LSST main survey will produce numerous transient alerts in a given night of observing, there are several factors that will minimize the impact of contamination on ToO observations. These include:

\begin{enumerate}

\item Previous studies have shown that the colors and timescales of kilonovae can be used to separate them from other transient populations \citep[see e.g.,][]{CowpBerger15,Doctor+17,Cowp+17b}. Kilonovae will appear redder and evolve more rapidly than other common transients such as supernovae.

\item Despite the high detection rate of LSST, the actual rate of new transients that appear during the narrow observing window and in the relatively small sky area will be low compared to a complete night of survey operations.

\item Previous imaging of the targeted sky area will be able to eliminate the presence of previously-existing transients that are older than \apx3 days, while continued monitoring will quickly eliminate longer-lived transients.

\item Rapidly evolving unclassified transients \citep[see e.g.,][]{Drout+14, Pursiainen+18} are potential contaminants, but their low rate ($\apx5\%$ of the SNe rate) means that the chance of one occurring in $\apx10$~deg$^2$ is small. 

\item Other rapidly evolving contaminants such as stellar flares can be common \citep[see e.g.,][]{Berger+13,Cowp+17b}, but will be bluer and shorter lived than kilonovae. Furthermore, in the case of stellar flares, the quiescent counterpart will be clearly visible in pre-existing imaging.

\item The effects of contamination can be further mitigated by additional post-processing techniques such as matching potential candidates against catalogs of probable host galaxies as determined from the ALV distance and localization information.
\end{enumerate}

\subsection{Final Considerations}
\label{sec:too_considerations}
To summarize, we have shown that effective and efficient ToO follow-up observations of GW signals from ALV can be conducted using LSST. We argue that observations in $g$ and $z$-bands with an integration time of $\apx180$~s each will be sufficient to robustly detect kilonovae. If we assume that LSST will require two sets of observations per 10 deg$^2$ of sky area to cover CCD gaps, then we estimate (including overheads) that LSST will require approximately 0.3 hours per epoch per 10 deg$^2$ observed. The total time investment then scales depending on choice of cadence (see \autoref{sec:too_early}). 

Using the inferred BNS merger rate from O2, ALV at design sensitivity will detect \apx50 mergers per year. Assuming that roughly a third\footnote{Accounting for sky position and Sun constraints} of those events are observable by LSST, a typical localization region is \apx20 deg$^2$ for the network (LIGO, Virgo, KAGRA, and potentially LIGO-India), and five epochs of observing time; then this represents \apx50 hours per year of LSST time spent on follow-up. This is $\apx1.5\%$ of the total survey time. The total sky area observed will be \apx 600 deg$^2$. Therefore, the total time and area impact of such follow-up efforts are minimal compared to the total survey time and area. The impact of main survey science goals could be further mitigated if observations taken are able to be incorporated back into the WFD survey.

\section{Discussion and Conclusion}
\label{sec:conc}
We presented detailed simulated LSST observations of kilonovae in two distinct modes. First, we investigated the ability of LSST to detect and characterize kilonovae over the course of the regular WFD survey with several choices of cadence. Second, we investigated the effectiveness of triggered ToO observations in response to GW triggers. Our key results can be summarized as follows:

\begin{enumerate}
\item The WFD survey has an average efficiency for kilonova detection of $\apx1-3\%$, with the WFDD cadence as the best performer. If we require the detection of a peak and color, then the efficiency drops to $\lesssim 0.5\%$. This efficiency is strongly a function of distance, peaking for nearby sources ($\lesssim 0.2$ Gpc), which is the typical detection distance for BNS mergers by the GW detector network. 
\item Using the calculated efficiency, we find that the WFD survey will yield $\approx 3-6$ kilonova detections per year, with the WFDD survey having the highest rate.  These numbers decrease by a factor of $2-5$ for events in which both the peak and color can be measured, with the WFDD survey having the smallest drop-off in efficiency.  The distance distribution for the small number of detected events peaks at $\apx0.2$~Gpc, comparable to the average detection distance for BNS mergers by a GW detector network.  Therefore, essentially all of the kilonova detected by LSST are likely to be detected in GW anyway.
\item We find that even the best kilonova light curves from the WFD survey will have insufficient data quality and light curve sampling to extract accurate physical parameters. Modeling the top $\apx10\%$ of detected kilonovae yield ejecta parameters that are poorly constrained and only accurate to within a factor of about $2-3$. These best-case events correspond to $3-6$ kilonova across the entire ten year survey.
\item Motivated by the poor performance of the WFD survey, and the fact that any detected kilonova are in any case within the GW network detection distance, we investigate LSST ToO follow-up of GW triggers. We find that such observations can be highly effective at detecting kilonovae, with a required exposure time of $\lesssim 300$ s in any filter. Specifically, just 2 hours per 10 deg$^2$ per event will lead to a detection efficiency of $\gtrsim 90\%$. We argue that observations in $gz$ bands will provide a useful minimal set, with a time sampling that leads to rapid characterization and discovery.  This minimal data set can of course be expanded on with additional filters and visits, but it represents a robust minimal time investment. In this context, LSST is most effective as a discovery instrument for the rapid identification of kilonovae candidates, which can then be handed off to other facilities for confirmation and further follow-up observations. 
\end{enumerate}

Ultimately, we have shown that the cadence of the LSST main survey is poorly matched to kilonova timescales, leading to just a few, poorly-characterized detections per year. On the other hand, a modest investment of at most $2\%$ of the WFD survey time can lead to robust detections of $\sim 200$ kilonovae during the ten year mission, with the resulting events being well characterized, and equally important, supported by GW information that will allow true multi-messenger studies of the events. It is important to note that kilonovae discovered before the start of LSST operations will improve our understanding of the range of model behaviors. While this does not impact our conclusions about the WFD survey, it will allow us to develop better ToO strategies. Nevertheless, LSST has the clear potential to be the premiere GW follow-up instrument for both current and future generations of GW detectors. 

\acknowledgements{We thank Rafaella Margutti and Brian Metzger for helpful discussions during the preparation of this manuscript. P.S.C. is grateful for support provided by NASA through the NASA Hubble Fellowship grant \#HST-HF2-51404.001-A awarded by the Space Telescope Science Institute, which is operated by the Association of Universities for Research in Astronomy, Inc., for NASA, under contract NAS 5-26555.}

\bibliographystyle{yahapj}
\bibliography{references}

\end{document}